\documentclass[11pt]{article} %
\usepackage{graphicx}
\usepackage{fancyhdr}
\usepackage{makeidx}
\usepackage{amssymb,amsmath}
\usepackage{physics}
\usepackage{upquote}
\usepackage{pdfpages}
\usepackage{xcolor}
\usepackage{enumerate}
\usepackage[compress,super,comma]{natbib}
\newcommand{\bd}{\begin{document}}
\newcommand{\ed}{\end{document}}
\newcommand{\bc}{\begin{center}}
\newcommand{\ec}{\end{center}}
\newcommand{\vs}{\vspace}
\newcommand{\hs}{\hspace}
\newcommand{\beq}{\begin{equation}}
\newcommand{\eeq}{\end{equation}}
\newcommand{\beqs}{\begin{eqn*}}
\newcommand{\eeqs}{\end{eqn*}}
\newcommand{\bq}{\begin{quote}}
\newcommand{\eq}{\end{quote}}
\newcommand{\lb}{\linebreak}
\newcommand{\mb}{\makebox}
\newcommand{\fb}{\framebox}
\newcommand{\mc}{\multicolumn}
\newcommand{\ben}{\begin{enumerate}}
\newcommand{\een}{\end{enumerate}}
\newcommand{\bit}{\begin{itemize}}
\newcommand{\eit}{\end{itemize}}
\newcommand{\ov}{\overline}
\newcommand{\un}{\underline}
\newcommand{\lt}{\left}
\newcommand{\rt}{\right}
\newcommand{\ba}{\begin{array}}
\newcommand{\ea}{\end{array}}
\newcommand{\beqa}{\begin{eqnarray}}
\newcommand{\eeqa}{\end{eqnarray}}
\newcommand{\beqas}{\begin{eqnarray*}}
\newcommand{\eeqas}{\end{eqnarray*}}
\newcommand{\bfg}{\begin{figure}}
\newcommand{\efg}{\end{figure}}
\newcommand{\pad}{\partial}
\newcommand{\nn}{\nonumber}
\newcommand{\la}{\leftarrow}
\newcommand{\ra}{\rightarrow}
\newcommand{\lgla}{\longleftarrow}
\newcommand{\lgra}{\longrightarrow}
\newcommand{\La}{\Leftarrow}
\newcommand{\Ra}{\Rightarrow}
\newcommand{\Lra}{\Leftrightarrow}
\newcommand{\Lgla}{\Longleftarrow}
\newcommand{\Lgra}{\Longrightarrow}
\renewcommand{\a}{\alpha}
\renewcommand{\b}{\beta}
\newcommand{\g}{\gamma}
\newcommand{\G}{\Gamma}
\renewcommand{\d}{\delta}
\newcommand{\D}{\Delta}
\newcommand{\e}{\epsilon}
\newcommand{\eps}{\epsilon}
\newcommand{\s}{\sigma}
\renewcommand{\l}{\lamda}
\newcommand{\m}{\mu}
\newcommand{\n}{\nu}
\renewcommand{\S}{\Sigma}
\newcommand{\p}{\pi}
\newcommand{\om}{\omega}
\newcommand{\Om}{\Omega}
\newcommand{\tri}{\triangle}
\newcommand{\ti}{\times}
\newcommand{\f}{\frac}
\newcommand{\ds}{\displaystyle}
\newcommand{\bm}[1]{\mb{{\boldmath $#1$}}}
\newcommand{\alter}[2]{\lt\{ \ba{ll}#1 \\ #2 \ea \rt.}
\newcommand{\alt}[4]{\lt\{ \ba{ll}#1 & \mb{if \, \,}#2 \\ #3 & \mb{}#4 \ea
    \rt.}
\newcommand{\altn}[4]{\lt\{ \ba{rl}#1 & \mb{if \, \,}#2 \\ #3 & \mb{}#4 \ea
    \rt.}
\newcommand{\altif}[4]{\lt\{ \ba{ll}#1 & \mb{if \, \,}#2 \\ #3 &
\mb{if \, \,}#4 \ea \rt.}
\newcommand{\altnif}[4]{\lt\{ \ba{rl}#1 & \mb{if \, \,}#2 \\ #3 &
\mb{if \, \,}#4 \ea \rt.}
\newcounter{algc}
\newcounter{romc}
\newcounter{Alphc}
\newcommand{\bl}{\begin{list}{{\it Step} ~\arabic{algc}~:} {\usecounter{algc}
                \setlength{\topsep}{0pt} \setlength{\itemsep}{0pt}}}
\newcommand{\el}{\end{list}}
\newcommand{\blr}{\begin{list}{~\roman{romc}~:} {\usecounter{romc}
                \setlength{\topsep}{0pt} \setlength{\itemsep}{0pt}}}
\newcommand{\elr}{\end{list}}
\newcommand{\bla}{\begin{list}{~\Alph{Alphc}~:} {\usecounter{Alphc}
                \setlength{\topsep}{0pt} \setlength{\itemsep}{0pt}}}
\newcommand{\ela}{\end{list}}
\newcommand{\tsup}{\textsuperscript}
\newcommand{\tsub}{\textsubscript}

\newtheorem{theorem}{Theorem}
\begin{document}
\title{Light emission from the layered metal 2H-TaSe\tsub{2} and its potential applications}
\author{Mehak Mahajan$^{1||}$, Sangeeth Kallatt$^{1||}$, Medha Dandu$^1$, Naresh Sharma$^2$, \\Shilpi Gupta$^2$ and Kausik Majumdar$^{1*}$\\
$^1$Department of Electrical Communication Engineering, \\Indian Institute of Science, Bangalore 560012, India\\
$^2$Department of Electrical Engineering, \\Indian Institute of Technology Kanpur, Kanpur 208016, India\\
$^{||}$ These authors contributed equally
\\$^*$Corresponding author, email: kausikm@iisc.ac.in}
\date{}
\maketitle
{\abstract Conventional metals, in general, do not exhibit strong photoluminescence. 2H-TaSe\tsub2 is a layered transition metal dichalcogenide that possesses metallic property with charge density wave characteristics. Here we show that 2H-TaSe\tsub2 exhibits a surprisingly strong optical absorption and photoluminescence resulting from inter-band transitions. We use this perfect combination of electrical and optical properties in several optoelectronic applications. We show a seven-fold enhancement in the photoluminescence intensity of otherwise weakly luminescent multi-layer MoS\tsub2 through non-radiative resonant energy transfer from TaSe\tsub2 transition dipoles. Using a combination of scanning photocurrent and time-resolved photoluminescence measurements, we also show that the hot electrons generated by light absorption in TaSe\tsub2 have a rather long lifetime unlike conventional metals, making TaSe\tsub2 an excellent hot electron injector. Finally, we show a vertical TaSe\tsub2/MoS\tsub2/graphene photodetector demonstrating a responsivity of $>10$ AW$^{-1}$ at $0.1$ MHz - one of the fastest reported photodetectors using MoS\tsub{2}.}

\newpage
\section{Introduction}
Adding new functionalities to devices is one of the important achievements of nano-materials and their heterojunctions. To this end, semiconducting two-dimensional (2D) transition metal dichalcogenides (TMDCs) have been extensively investigated for a plethora of nanoelectronic and optoelectronic applications \cite{Lee2014,Withers2015a,Shim2017}. In general, obtaining a low resistance electrical contact without destroying the high quality of ultra-thin 2D active layers remains an outstanding challenge for 2D materials. Recently, vertical electronic and optoelectronic devices \cite{Lee2014, Massicotte2015, Kallatt2018, Withers2015} using layered materials are becoming increasingly popular as they exploit carrier transport through these nanometer-thick active layers - something that is difficult to achieve in a lithography limited conventional planar structure. Maintaining the pristine nature of the layers underneath the contact becomes even more crucial in the context of these vertical devices where the contact physically sits right on top of the active layers.
\\
\\
In this context, metallic TMDCs would be an ideal choice for contact material. However, metallic TMDCs remain a less explored class of materials. 2H-TaSe\tsub2 is one such layered TMDC that shows metallic properties \cite{Wilson1974, Lee1970, Naito1982, Neal2014} and would be promising as a contact material for nanoelectronic devices. 2H-TaSe\tsub2 exhibits simultaneously strong electron-phonon coupling and spin-orbit coupling \cite{Rossnagel2007}, which play a crucial role in charge density wave (CDW) driven transport properties. It exhibits highest CDW transition temperature among TMDCs exhibiting 2H symmetry. The resulting bandstructure shows a narrow band around Fermi energy with uniquely structured Fermi surfaces. Interestingly, there is a large gap above and below the bands around the Fermi energy \cite{Laverock2013, Renteria2014, Tsoutsou2016, Li2018} in the normal metal phase and is expected to exhibit interesting optical properties, which remain largely unexplored. The perfect combination of excellent electrical transport and optical properties, coupled with the possibility of wafer-scale growth \cite{shi2018chemical},  makes 2H-TaSe\tsub2 a unique material for flexible, ``all-2D" optoelectronic applications.
\\
\\
Motivated by this, in this work, we explore 2H-TaSe\tsub2 as a promising multi-purpose optoelectronic material. We show that this material exhibits strong optical absorption and photoluminescence, while maintaining metallic conductivity behavior. These properties have been explored here to demonstrate versatile device applications of 2H-TaSe\tsub2, for example, (1) as a donor layer for non-radiative resonant energy transfer (NRET) \cite{Kozawa2016,Guzelturk2016,Deshmukh2018,dandu2019} to other material, (2) as an efficient hot electron injector, (3) as a contact material for planar and vertical devices, and finally (4) as a dual-purpose layer, namely as light absorber as well as photo-carrier collector, in a sensitive, high speed vertical photodetector.
\\
\\
\section{Results}
We obtain TaSe\tsub2 flakes on Si substrate covered with thermally grown $285$ nm SiO\tsub2 using micromechanical exfoliation from bulk crystals (procured from 2D Semiconductors). The bulk crystals are characterized using scanning electron micrograph (SEM), X-ray diffraction (XRD), and energy-dispersive X-ray spectroscopy (EDS) (see \textbf{Supplementary Note 1}). We  use temperature dependent resistivity measurement and Raman spectroscopy to determine the polytype of the TaSe\tsub2 samples we use in this work. 1T-TaSe\tsub2 has an octahedral coordination and exhibits a normal metal to incommensurate, and incommensurate to commensurate CDW phase transition at $600$ K and $473$ K, respectively, with the latter being associated with a sharp discontinuity in resistivity \cite{Wilson1974,Wilson1975,Inada1980}. On the other hand, the 2H-TaSe\tsub2 polytype exhibits a trigonal prismatic structure (see Figure 1a) and remains normal metal above $123$ K. Upon cooling below $123$ K, it undergoes a CDW phase transition from metallic to incommensurate phase. On further cooling, it undergoes a first order lock in transition to commensurate $3\times 3$ ordered transition around $90$ K \cite{Rossnagel2007, Li2018}. Unlike the 1T polytype, 2H-TaSe\tsub2 does not exhibit any discontinuity in resistivity. The current-voltage characteristics of a $40$ nm thick flake is shown in Figure 1b at different temperatures ($T$). The normal metal to incommensurate CDW phase transition shows a sharp change in the slope in the resistivity (extracted from Figure 1b) versus $T$ plot \cite{Lee1970, Naito1982, Neal2014}, as shown in Figure 1c around $120$ K. Above this temperature, 2H-TaSe\tsub2 resistivity follows a $AT+B$ functional form with $A$ and $B$ being constants, which is different from the $AT$ law a conventional metal obeys. The term $B$ arises from impurity scattering due to local CDW fluctuation \cite{Naito1982}. Below the transition temperature $T_0$, the resistivity drops as $T^2$ up to $T \sim 30$ K, and as $T^5$ for $T < 30$ K \cite{Naito1982}.
\\\\
2H-TaSe\tsub2 also exhibits conspicuous Raman active A$_{\textrm{1g}}$, E$_{\textrm{1g}}$ and E$_{\textrm{2g}}$ modes \cite{Neal2014,Hajiyev2013,Yan2015}, as illustrated in Figure 1d. The A$_{\textrm{1g}}$ peak corresponds to the out of plane vibration mode at a wave number of  $236$ cm\tsup{-1}, the E$_{\textrm{2g}}$ peak corresponds to in-plane vibrational mode at $210$ cm\tsup{-1} and a broad E$_{\textrm{1g}}$ peak at $140$ cm\tsup{-1} result from two-phonon processes. In \textbf{Supplementary Note 2}, we show the X-ray photoelectron spectrum of Ta core levels (4f\tsub{5/2} and 4f\tsub{7/2}), both before and after an \emph{in-situ} sputter etch step. The peak positions correlate well with TaSe\tsub2 \cite{luo2016differences}. Note that, in the pre-etch measurement, while we do not observe any distinct tantalum oxide peak, after curve fitting, a higher energy peak is found which shows presence of an ultra-thin surface oxide layer. This self-limiting layer acts as a protective layer to further oxidation. We provide a detailed short term and long term stability analysis of the TaSe\tsub2 flakes in \textbf{Supplementary Note 3} using a combination of optical imaging, atomic force microscopy and Raman spectroscopy.
\\\\
\subsection{Photoluminescence from metallic TaSe\tsub2}
In spite of showing metal-like conductivity behavior, the 2H-TaSe\tsub2 flakes exhibit surprisingly strong photoluminescence (PL). We acquire PL spectra from TaSe\tsub2 flakes of varying thickness using continuous wave laser of wavelength $532$ nm at different temperatures from $3.3$ K to $300$ K where the sample chamber is kept at a high vacuum level of less than $10^{-7}$ bar. The laser power is kept below $150$ $\mu$W during the PL experiment. The obtained spectra from flakes of two different thickness values (monolayer and $50$ nm thick multi-layer) are shown in Figure 2a-b. Multiple features can be readily identified in the spectra: (1) Both the films exhibit a broad prominent peak around $2.0$ eV - $2.1$ eV. (2) No such strong luminescence is observed when an excitation with lower energy photon ($633$ nm) is used. (3) The peak position blue shifts by about $30$-$50$ meV depending on the thickness as the temperature is lowered from $300$ K to $3.3$ K. (4) Upon cooling below $123$ K, and further below $90$ K, the bands of 2H-TaSe\tsub2 are known to get significantly distorted to a ``spaghetti-like" structure due to reconstruction \cite{Rossnagel2007}. However, we do not observe any strong signature of such reconstruction in the photoluminescence. (5) For the thick film, a weak peak is identified around $1.75$ eV (indicated by red arrow in Fig 2b), particularly at higher temperature, and this is negligible in the monolayer. Figure 2c summarizes the thickness dependence of the peak position, measured over a large number of samples at $300$ K, suggesting a modulation of about $100$ meV can be achieved in the emission energy by changing the thickness of the flake. It is worth noting that photoluminescence has been reported in the past from films of tantalum based compounds which have been intentionally oxidized by exposing to relatively higher power laser beam \cite{castellanos2013fast, coronado2013nanofabrication, cartamil2015high}. However, in our experiment, the origin of such strong PL from laser induced oxidation can be ruled out since the measurement is performed in high vacuum condition, and the laser power is also kept about an order of magnitude lower. In \textbf{Supplementary Note 4}, we provide a full spectrum of the PL showing the Raman lines close to the laser excitation, which suggests absence of any oxide Raman peak. Also, the strong dependence of the photoluminescence peak positions on the thickness of the TaSe\tsub2 films clearly suggests that the observed luminescence is not resulting from any surface-oxide layer. To further establish the point that any ambience induced surface oxidation (during exfoliation) does not give rise to the observed photoluminescence, we perform exfoliation followed by Poly(methyl methacrylate) (PMMA) spin coating inside a glovebox with N\tsub2 atmosphere. The coated samples exhibit similar photoluminescence and Raman characteristics, including a shift in the photoluminescence peak with change in flake thickness, as explained in \textbf{Supplementary Note 5}.
\\\\
The general features of the light absorption and PL spectra can be qualitatively understood from the reported electronic structure of 2H-TaSe\tsub2, both using density functional calculations \cite{Laverock2013,Renteria2014,Yan2015,Kuchinskii2012}, tight-binding calculations \cite{Rossnagel2007,Laverock2013,Smith1985} and angle resolved photoemission spectroscopy (ARPES) measurements \cite{Laverock2013,Tsoutsou2016,Li2018,Smith1985,Jakovidis1992}. With an excitation energy of $2.33$ eV, two possible direct optical transitions can be identified, as illustrated in the  schematic bandstructure of a few-layer-thick 2H-TaSe\tsub2 (in the normal metal state, without any reconstruction) in Figure 2d. Like other layered CDW materials, 2H-TaSe\tsub2 has a narrow band around the Fermi energy across the Brillouin zone. There are also a set of bands below and above it which are well separated in energy. Transition 1 is a direct inter-band transition from the bottom band to the band above the Fermi energy at $K$ point. The photo-carriers (both electrons and holes) thus generated are scattered in the k-space in a very short time scale. This process is similar to gold and copper \cite{Mooradian1969}, and results in a weak luminescence. On the other hand, transition 2 (in between $\Gamma$ and $K$) causes a strong optical absorption due to the parallel nature of the band pair. The photo-generated electrons from the upper band relax to the $K$ and $\Gamma$ valley through indirect transitions with the assistance of phonons - resulting in luminescence.
\\\\
The full width at half maximum (FWHM) of the peaks is $\sim 250$-$300$ meV and is significantly larger compared with typical excitonic resonance observed in monolayers of semiconducting TMDCs (like MoS\tsub2, MoSe\tsub2, WS\tsub2, and WSe\tsub2). The linewidth remains large irrespective of the thickness and sample temperature, even down to $3.3$ K and can be understood from the indirect nature of carrier relaxation, as well as strong scattering from the free carriers. We also did not observe any sharp excitonic feature in differential reflectance spectra of the flakes, as shown in \textbf{Supplementary Note 6}.
\\
\\
In order to probe the lifetime of the photo-excited carriers in TaSe\tsub2, we perform time-resolved photoluminescence (TRPL) measurements. The measurement details and the results are summarized in \textbf{Supplementary Note 7}. Supplementary Figure 10d shows the statistical distribution of the lifetimes extracted after deconvolving the observed data with instrument response function (IRF) across all measurements (measured over seven different flakes, several spots on each flake). The dotted curves in Supplementary Figure 10d represent Gaussian fit to the distributions. The mean values of extracted lifetimes are 6.8 ps for 45 $\mu$W and 2.5 ps for 140 $\mu$W of incident power at 405 nm excitation wavelength. The distributions are found to be quite tight, with a standard deviation of 1.4 ps for 45 $\mu$W and 0.9 ps for 140 $\mu$W of excitation power. We, however, note that there may be some inaccuracy in these extracted lifetime values due to the closeness of the measured data with IRF. These lifetime values for the photo-carriers are significantly longer than, for example, what one would get in conventional metallic systems - a key result that is exploited later for efficient hot electron injection from TaSe\tsub2 to other semiconducting TMDC.
\\
\\
\subsection{Resonant energy transfer between TaSe\tsub2 and MoS\tsub2}
The broad emission linewidth from TaSe\tsub2 has a large spectral overlap with the excitonic absorption of other semiconducting TMDC materials like $A_{\textrm{1s}}$ exciton of MoS\tsub2, as schematically shown in Figure 3a. We exploit this feature to transfer energy from TaSe\tsub2 to other semiconducting TMDC layers using dipole-dipole coupling through NRET process \cite{Kozawa2016,Guzelturk2016,Deshmukh2018}. Due to layered structure, the transition dipoles formed in both MoS\tsub2 and TaSe\tsub2 are quasi-two-dimensional in nature. The dipole-dipole interaction between these two vertically stacked materials is facilitated due to such in-plane orientation of the transition dipoles, which allow matching of the center of mass momentum \cite{Lyo2000}, as schematically depicted in Figure 3b. We prepare multiple TaSe\tsub2 (top)/multi-layer MoS\tsub2 (bottom) vertical heterojunction stacks on Si/SiO\tsub2 substrate (see Methods). Figure 3c shows the PL spectrum from the part of the TaSe\tsub2 flake which is not on the junction. The data can be fitted with three different Voigt peaks, as shown by the dashed lines.  We use a multi-layer MoS\tsub2 to ensure that the luminescence from the MoS\tsub2 flake alone is extremely weak.  Figure 3d shows the weak MoS\tsub2 peak around $1.85$ eV which corresponds to the direct band gap transition at the $K (K^\prime)$ point of the Brillouin zone. While plotting, the MoS\tsub2 peak intensity in multiplied by 5.
\\\\
Figure 3e shows a seven-fold enhancement in the MoS\tsub2 PL intensity (indicated by the arrow) in the TaSe\tsub2/MoS\tsub2 junction. The resulting photoluminescence spectrum from the junction can now be fitted with four Voigt peaks, three from TaSe\tsub2 (in dashed traces) and the fourth one representing MoS\tsub2 emission (in solid black trace). At the junction, while the MoS\tsub2 peak is enhanced, the individual dashed traces clearly indicate the suppression of the strength of the TaSe\tsub2 emission at the emission energy of MoS\tsub2, justifying energy transfer from TaSe\tsub2 to $A_{\textrm{1s}}$ exciton. The underlying mechanism is explained schematically in Figure 3f where the rates of the individual processes are indicated. In multi-layer MoS\tsub2, the direct peak is weak due to fast (on the order of $100$ fs \cite{Bertoni2016}) transfer of carriers to lower energy indirect valleys. The photo-excited carriers generated by an excitation with $532$ nm (i.e. $2.33$ eV) photon - that is off-resonant with the 1s peak - transfers to the lower valleys rather than forming an exciton in the $K$ and $K^\prime$ valleys.
\\\\
According to the experimental data, photoluminescence of the MoS\tsub2 A$_{\textrm{1s}}$ peak gets enhanced at the MoS\tsub2/TaSe\tsub2 junction suggesting a gain channel for exciton density. MoS$_2$ gains excitons at the $K$ ($K^\prime$) valley at a Forster NRET rate ($\mathit{\Gamma}_{\textrm{ET}}^{\textrm{T-M}}$) from the energy states in TaSe\tsub2 that are resonant to the exciton state of MoS\tsub2. The MoS\tsub2 excitons decay through multiple parallel channels, namely the radiative recombination ($\mathit{\Gamma}_\textrm{r}$) and  non-radiative processes including scattering to lower energy indirect valley in MoS\tsub2 ($\mathit{\Gamma}_{\textrm{in}}$), as well as Dexter ($\mathit{\Gamma}_{\textrm{CT}}^{\textrm{M-T}}$) and Forster ($\mathit{\Gamma}_{\textrm{ET}}^{\textrm{M-T}}$) process induced exciton back transfer to TaSe\tsub2. With this understanding, we form the rate equations for exciton density in MoS$_2$ ($N_{\textrm{ex}}^{\textrm{M,jun}}$) and free e-h pair density in TaSe$_2$ ($N_{\textrm{e-h}}^{\textrm{T,jun}}$) on junction, as explained in \textbf{Supplementary Note 8}.
%can be expressed as:
%\begin{equation}\label{eq:m}
%\frac{dN_{ex}^{M,jun}}{dt}=\Gamma_{ex}N_{e-h}^{M,jun}-(\Gamma_{d}+\Gamma_{CT}^{M-T}+\Gamma_{ET}^{M-T})N_{ex}^{M,jun}+\Gamma_{ET}^{T-M}N_{e-h}^{T,jun}
%\end{equation}
%
%\begin{equation}\label{eq:mt}
%\frac{dN_{e-h}^{T,jun}}{dt}= G_{f}^{T,jun}-(\Gamma_{s}+\Gamma_{ET}^{T-M})N_{e-h}^{T,jun}+(\Gamma_{CT}^{M-T}+\Gamma_{ET}^{M-T})N_{ex}^{M,jun}
%\end{equation}
%where G$_{f}^{T,jun}$ is the effective generation rate of e-h pairs in the energy state of TaSe$_2$ that is resonant to the exciton state in MoS$_2$ and $\Gamma_{s}$ is the scattering rate of these e-h pairs from this state to other non-resonant energy states in TaSe$_2$. $\Gamma_d=\Gamma_r+\Gamma_{in}$ is the combined radiative and non-radiative decay rate of excitons within MoS\tsub2 and is assumed to be similar on SiO\tsub2 and TaSe\tsub2.
%\\
%\\
%With the observation that the photoluminescence of A$_{1s}$ peak of MoS\tsub2 on TaSe\tsub2 enhances significantly relative to that of MoS\tsub2 on SiO\tsub2, the first term in equation \ref{eq:m} representing direct absorption by MoS\tsub2 can be dropped. The steady state photoluminescene intensity of A$_{1s}$ peak is then given by
%\begin{equation}\label{eq:I}
%   I_{M}^{jun}= \Gamma_{r}N_{ex}^{M,jun}= \frac{G_{e-h}^{T,jun}\Gamma_{ET}^{T-M}\Big[\dfrac{\Gamma_{r}}{\Gamma_{d}}\Big]}{\Gamma_{s}\Big[1+\dfrac{\Gamma_{CT}^{M-T}+\Gamma_{ET}^{M-T}}{\Gamma_{d}}\Big]+\Gamma_{ET}^{T-M}}
%\end{equation}
%\\
The PL enhancement factor ($\mathit{\alpha}$) is defined to be the ratio of the steady state PL intensity of MoS\tsub2 on TaSe\tsub2 to that of on SiO\tsub2 and turns out to be
\begin{equation}\label{eq:Igain}
  \mathit{\alpha}= \dfrac{I_{\textrm{M}}^{\textrm{jun}}}{I_\textrm{M}}= \dfrac{G_{\textrm{e-h}}^{\textrm{T,jun}}}{G_{\textrm{ex}}^\textrm{M}}\Bigg\{\dfrac{\mathit{\Gamma}_{\textrm{ET}}^{\textrm{T-M}}}{\mathit{\Gamma}_{\textrm{s}}\Big[1+\dfrac{\mathit{\Gamma}_{\textrm{CT}}^{\textrm{M-T}}+\mathit{\Gamma}_{\textrm{ET}}^{\textrm{M-T}}}{\mathit{\Gamma}_{\textrm{d}}}\Big]+\mathit{\Gamma}_{\textrm{ET}}^{\textrm{T-M}}}\Bigg\}
\end{equation}
where relevant variables are defined in \textbf{Supplementary Note 8}. Equation \ref{eq:Igain} suggests that the enhancement factor increases as $\mathit{\Gamma}_\textrm{s}$ decreases due to improved energy transfer with a maximum attainable value of $\mathit{\alpha}_{\textrm{max}} = \dfrac{G_{\textrm{e-h}}^{\textrm{T,jun}}}{G_{\textrm{ex}}^{\textrm{M}}}$. Such an efficient resonant energy transfer process is lucrative for a variety of optoelectronic applications, including electroluminescence \cite{Puchert2017}.
\\\\
\subsection{TaSe\tsub2 as an efficient hot carrier injector}
The strong light absorption by TaSe\tsub2 makes it a promising material for hot carrier injection to other semiconducting materials. We fabricate MoS\tsub2/TaSe\tsub2 heterojunction device with Ni/Au as contact pads, as schematically shown in Figure 4a (see Methods). A laser beam of wavelength $532$ nm is scanned from one contact to the other (along the arrow in the optical image in Figure 4b), and we measure the corresponding photocurrent (see Methods) at different external bias conditions. Figure 4c shows the current-voltage characteristics of the heterojunction for a few selected positions of the laser spot, as indicated by numbers. The dark current from the device is represented by the black line. The spatial dependence of the photocurrent is illustrated in Figure 4d at $V_{\textrm{ds}}=-0.1$ V by plotting the photocurrent as a function of the laser position.
\\\\
When the laser spot is on the TaSe\tsub2 flake with no MoS\tsub2 underneath (points 5 and 6), we observe a strong photoresponse at a non-zero external bias. This strong photocurrent away from the junction is surprising when compared to other reported metal-TMDC junction devices \cite{Hong2015}. A weak photocurrent on the metal portion away from the junction has been attributed earlier to photo-thermoelectric effect due to laser spot induced local heating. However, in the present case, such a photo-thermoelectric effect can be ignored as the photocurrent strongly depends on the applied external bias, and is negligible under zero bias. This is further verified by noting that the photocurrent is extremely weak when a $785$ nm excitation is used (Figure 4e) which does not allow any inter-band transition in TaSe\tsub2. The strong photocurrent (with 532 nm) with the laser spot on TaSe\tsub2 suggests that the photo-electrons, generated by the inter-band transition in TaSe\tsub2, exhibit relatively longer recombination lifetime compared to other metals. This is in good agreement with the TRPL measurement discussed earlier. The long lifetime allows us to collect the photoelectrons at the contact even when the separation between the laser spot and the metal contact is several micrometers (for example, point 6 in Figure 4b), which is few orders of magnitude longer than conventional metals. This makes TaSe\tsub2 particularly attractive material as a source for efficient hot carrier injection techniques.
\\\\
When the excitation wavelength is 532 nm, the device exhibits a large open circuit voltage ($V_{\textrm{oc}}$) and short circuit current ($I_{\textrm{sc}}$) which strongly depend on the position of the laser spot, as explained in Figure 4c. The asymmetry between Au/MoS\tsub2 and TaSe\tsub2/MoS\tsub2 junctions provides a strong built-in field in the device. $I_{\textrm{sc}}$ is maximum when the laser spot is directly on the MoS\tsub2 channel, and this leads to a $V_{\textrm{oc}}$ of $0.1$ V. $V_{\textrm{oc}}$ and $I_{\textrm{sc}}$ are zero on the other hand when the laser is on TaSe\tsub2 as there is no built-in field in TaSe\tsub2 owing to its high electrical conductivity. In between, upon moving the laser spot away from the MoS\tsub2 channel towards right along the arrow in Figure 4b, the $V_{\textrm{oc}}$ and $I_{\textrm{sc}}$ are gradually suppressed. Using transfer length theory \cite{Schroder, Majumdar2013} of metal-semiconductor contacts, the built-in field dies rapidly in the lateral direction in the MoS\tsub2 portion underneath TaSe\tsub2 (see top panel of Figure 4d), resulting in a gradual suppression of the zero-bias photocurrent as the laser spot is moved away from the MoS\tsub2 channel.
\\\\
\subsection{Graphene/MoS\tsub2/TaSe\tsub2 vertical junction}
Inspired by the strong built-in electric field in the lateral heterojunction device in Figure 4, we next fabricate a vertical heterojunction 2H-TaSe\tsub2/2H-MoS\tsub2/1L-graphene (see Methods), as schematically shown in Figure 5a. Two-dimensional layered materials have recently attracted a lot of attention as active material in photodetector applications \cite{koppens2014photodetectors,li2017graphene,mak2016photonics}. Vertical heterojunction photodetectors based on ultra-thin two-dimensional layered materials provide a number of advantages \cite {Kallatt2018, Murali2018} compared to lateral structures, namely (1) ultra-short carrier transit time due to nanometer separated electrodes, which is difficult to achieve otherwise in a lithography limited planar structure; (2) large built-in vertical field; (3) suppression of series resistance due to improved carrier collection efficiency that is not limited by transfer length; and (4) repeatable characteristics where encapsulated photoactive layer does not degrade from ambience effects and screened from substrate induced traps.
\\\\
The difference in the work function between TaSe\tsub2 and monolayer (1L) graphene provides a strong built-in field across the few-layer MoS\tsub2 film which is sandwiched between the two. The structure utilizes the excellent electrical contact between TaSe\tsub2 and MoS\tsub2, while maintaining the pristine nature of MoS\tsub2 at the junction which is critical for the operation of such vertical heterojunction. Monolayer graphene at the bottom serves two purposes. First, it acts like the bottom electrode for carrier collection. Second, using the weak screening through graphene, a back-gate can be used to modulate the chemical potential of the MoS\tsub2 sandwiched layer.
\\\\
The optical image of the fabricated heterojunction device is shown in Figure 5b. The photocurrent ($I_{\textrm{ph}}$) from the device is obtained as $I_{\textrm{ph}}=I-I_{\textrm{dark}}$ where $I$ is the measured current in presence of light and $I_{\textrm{dark}}$ is the dark current. Figure 5c illustrates the variation of $I_{\textrm{ph}}$ as a function of external bias when excited with a diffraction limited laser spot (diameter $\sim 1.3$ $\mu$m) of wavelength $532$ nm. The vertical built-in field in the device manifests as a nonzero photocurrent at zero external bias. Three possible mechanisms that can contribute to the photocurrent are schematically depicted in Figure 5d-f. The left and right panels represent the state transition diagrams and the electronic band diagrams, respectively. The first process is explained in Figure 5d. NRET driven energy transfer due to dipole-dipole interaction between TaSe\tsub2 and MoS\tsub2 results in a creation of exciton in MoS\tsub2 due to light absorption in TaSe\tsub2. The exciton thus created in MoS\tsub2 is dissociated at a fast time scale due to scattering to lower energy indirect valleys, creating free electron and hole at the band edges. These electrons and holes are in turn separated by the built-in field, giving rise to photocurrent. Such an energy transfer driven photocurrent generation is a unique way of photodetection, and can find multiple applications in van der Waals heterojunctions. Second, the light absorbed by TaSe\tsub2 creates hot electrons, which are injected into MoS\tsub2. The built-in field in MoS\tsub2 carries these electrons towards graphene, resulting in photocurrent, as shown in Figure 5e. Third, the electron-hole pairs generated by the photons directly absorbed in MoS\tsub2 also result in population in the band edge indirect valleys, causing photocurrent, as explained in Figure 5f.
\\\\
In order to segregate the effect of the above three mechanisms, the photocurrent is measured with $785$ nm excitation (which is not absorbed by TaSe\tsub2), and the current-voltage characteristics are summarized in the inset of Figure 5c. We notice that the photocurrent at zero external bias is completely suppressed. This indicates that the zero-bias photocurrent for $532$ nm excitation is primarily governed by TaSe\tsub2 induced hot carrier transfer and dipole-dipole interaction driven NRET process.
\\\\
The photocurrent is plotted as a function of the incident optical power ($P_{\textrm{op}}$) of 532 nm excitation in Figure 6a. The data can be fitted with a power law: $I_{\textrm{ph}}\propto P_{\textrm{op}}^{0.52}$, which indicates that the traps due to the gap states in MoS\tsub2 play an important role in determining $I_{\textrm{ph}}$. The presence of gap states in MoS\tsub2 (see Figure 6b) is also identified by the back gate modulation of the electrical conductivity of the device. Figure 6c shows the transfer characteristics [drain current ($I_\textrm{d}$) versus gate voltage ($V_\textrm{g}$)] where $V_\textrm{g}$ is applied at the back side. Upon cooling the device to 77 K, we obtain an on-off ratio of $\sim 6.6\times10^3$, which reduces to $\sim 45$ at $280$ K. The on-off ratio at $280$ K suffers from tunneling of carriers from bottom graphene electrode to top TaSe\tsub2 electrode though the bandgap states in MoS\tsub2 resulting from band tail states as well as deep traps due to defects \cite{Furchi2014}. The tunneling effect manifests in a particularly strong manner for large negative $V_\textrm{g}$, where the gate voltage dependence on the drain current is completely absent due to the tunneling leakage. When the device is cooled down to $77$ K, the trap assisted tunneling can be reduced significantly. The output characteristics [$I_\textrm{d}$ versus drain voltage ($V_\textrm{d}$)] of the vertical device at $77$ K are displayed in Figure 6d.
\\\\
Figure 6e shows the obtained responsivity $(R=\frac{I_{\textrm{ph}}}{P_{\textrm{op}}})$ from the device at different incident optical powers. The responsivity is $>10$ at the minimum power used, which suggests an internal gain in the device. The different photo-carrier generation mechanisms described earlier also populate the trap states closer to the valence band of MoS\tsub2 with holes. This in turn results in a gain mechanism \cite{Murali2018,Furchi2014}. Two possible mechanisms, namely photoconductive gain and photogating effect, provide gain in the system. Photo-generated holes trapped in the gap states of the MoS\tsub2 film cause a charge imbalance forcing reinjection of electrons multiple times before the trapped hole recombines or tunnels out of MoS\tsub2 - providing photoconductive gain. Electron trapping is neglected as the Fermi level in MoS\tsub2 lies close to the conduction band edge. Since the photoconductive gain is given by $\frac{\tau}{\tau_{\textrm{tr}}}$, where $\tau$ is the lifetime of the hole in the trap state, the vertical structure is advantageous due to its short transit time ($\tau_{\textrm{tr}}$) of electrons through the thin MoS\tsub2 layer \cite{Kallatt2018}. On the other hand, the trapped holes suppress the barrier for carrier injection to MoS\tsub2 from vertical electrodes, providing a photogating effect.
\\\\
The heterostructure device is further characterized for the speed of operation using a laser which is intensity modulated with varying frequency. The photovoltage generated by the device is measured across a digital storage oscilloscope which is terminated with $10$ M$\Omega$ impedance. Figure 6f shows the measured photovoltage in response to a $0.1$ MHz modulation frequency under zero external bias operation, which is one of the fastest reported photodetectors using TMDC material. The short transit time of the photo-electrons helps to obtain the high speed of operation of the detector, while maintaining the gain. In \textbf{Supplementary Table 1}, we compare the performance of the photodetector presented in this work with existing 2D material based detectors (having response time less than 100 ms).
\\
\\
\section{Discussion}
In summary, we demonstrate the versatility of 2H-TaSe\tsub2 as a two-dimensional layered CDW material, which exhibits metal-like characteristics, coupled with strong light absorption and emission due to energy separated bands. The narrow band around the Fermi energy results in giant metallic conductivity with temperature dependent CDW characteristics. This makes TaSe\tsub2 an ideal candidate for a 2D-2D van der Waals contact material which can maintain the pristine nature of the underneath 2D layered semiconductor. On the other hand, the well separated bands lying above and below the Fermi energy result in strong inter-band optical absorption, strong photoluminescence and a long lifetime of photoelectrons generating large photocurrent. The spectral overlap between emission of TaSe\tsub2 and excitonic absorption of other TMDC semiconductors like MoS\tsub2 results in a unique in-plane momentum matched 2D-2D combination of donor-acceptor pair causing a giant enhancement of photoluminescence in multi-layer MoS\tsub2 due to non-radiative resonant energy transfer. TaSe\tsub2 is further utilized as a dual purpose layer in a vertical photodetection device to obtain high speed and gain simultaneously. The large separation of the bands also make TaSe\tsub2 an interesting material for low-loss plasmonic applications \cite{Khurgin2010, Gjerding2017}. The finding will ignite multiple device applications using 2H-TaSe\tsub2 in the nano-opto-electronic domain.
\section{Methods}
\subsection{Heterojunction device fabrication}
For both planar and vertical devices, the 2D material stack is fabricated by using dry transfer technique on a highly doped Si substrate coated with $285$ nm thick thermally grown oxide. This stack is heated on a hot plate at $80^\circ$ C for 2 minutes for improved adhesion between layers. Devices are fabricated using standard nano-fabrication methods. The substrate is spin coated with PMMA $950$C$3$ and baked on a hot plate at $180^\circ$ C for $2$ minutes. This is followed by electron beam lithography with an acceleration voltage of $20$ KV, an electron beam current of $300$ pA, and an electron beam dose of $200$ $\mu$Ccm$^{-2}$. Patterns were developed using MIBK:IPA solution in the ratio $1:3$. Later samples are washed in isopropyl alcohol and dried in N$_2$ blow. $10$ nm Ni / $50$ nm Au films are deposited by using magnetron sputtering with the help of Ar plasma at $3\times10^{-3}$ mBar. Metal lift-off was done by dipping the substrate in acetone for $15$ minutes, followed by washing in isopropyl alcohol and $N_2$ drying. The bottom side of the wafer is etched by dilute HF solution and silver film is deposited as the bottom gate contact.\\
\subsection{Photocurrent measurements}
Devices are probed by micromanipulators, connected to a  Keithley 2636B and Agilent B1500 device analyzers for electrical measurements. Diffraction limited laser spots of wavelength 532 nm and 785 nm are used for photocurrent measurements through a $\times$50 objective with a numerical aperture of 0.5. The device is kept on a motorized stage to obtain spatial mapping of photocurrent. For low temperature electrical measurements, a liquid nitrogen cooled stage is used. High speed measurements are performed using a 532 nm laser beam. The intensity of the laser beam is modulated at the required frequency, up to 0.1 MHz and the photovoltage measured using a digital storage oscilloscope which is terminated by 10 M$\Omega$.
\section*{Data availability}
	The datasets generated and analyzed during the current study are available from the corresponding author on reasonable request.
\bibliographystyle{naturemag}
\bibliography{TSM}
\section*{Acknowledgements}
This work was partially supported in part by a grant under Indian Space Research Organization (ISRO), by the grants under Ramanujan Fellowship, Early Career Award, and Nano Mission from the Department of Science and Technology (DST), and by a grant from MHRD, MeitY and DST Nano Mission through NNetRA. S.G. acknowledges funding support from DST (SR/FST/ETII-072/2016) and SERB (SB/S2/RJN-134/2014, EMR/2016/007113).
\section*{Contributions}
K.M. designed the experiment. M.M. and S.K. prepared the samples, fabricated the devices and performed the measurements. N.S. and S.G. performed the time-resolved photoluminescence measurement. M.D. contributed to the theoretical analysis of the energy transfer mechanism. All authors contributed to the data analysis, discussion of results, and writing of the paper. M.M. and S.K. contributed equally to this work.
\section*{Competing interests}
	The Authors declare no competing interests.
\section*{Corresponding author}
Correspondence to K. Majumdar.
\newpage
\begin{figure}[!hbt]
\centering
%\vs{-0.1in}
%\hs{-1in}
\includegraphics[scale=0.5]{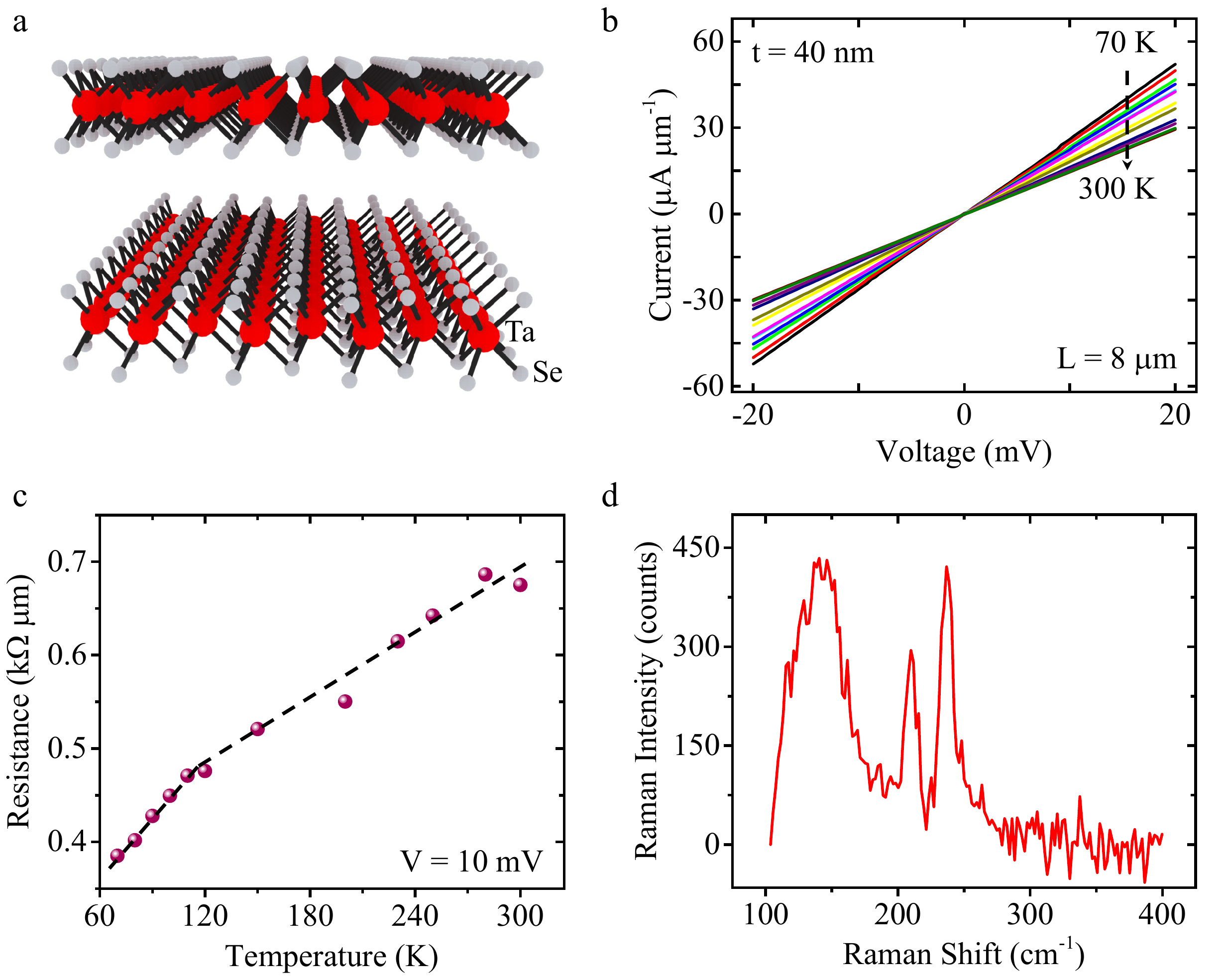}
%\vspace{-1.8in}
\caption{\textbf{Raman and electrical characterization of 2H-TaSe\tsub2.}(a) Schematic representation of crystal structure of 2H-TaSe\tsub2. (b) Current-voltage characteristics of a two-probe device using multi-layer 2H-TaSe\tsub2 as the channel at different temperatures. (c) Variation of resistance (scaled by the width of the flake) with temperature as extracted from (b). (d) Raman shift in multi-layer 2H-TaSe\tsub2 at room temperature using a 532 nm laser.}\label{fig:fig1}
\end{figure}
\newpage
\begin{figure}[!hbt]
\centering
%\vs{-0.1in}
%\hs{-1in}
\includegraphics[scale=0.5]{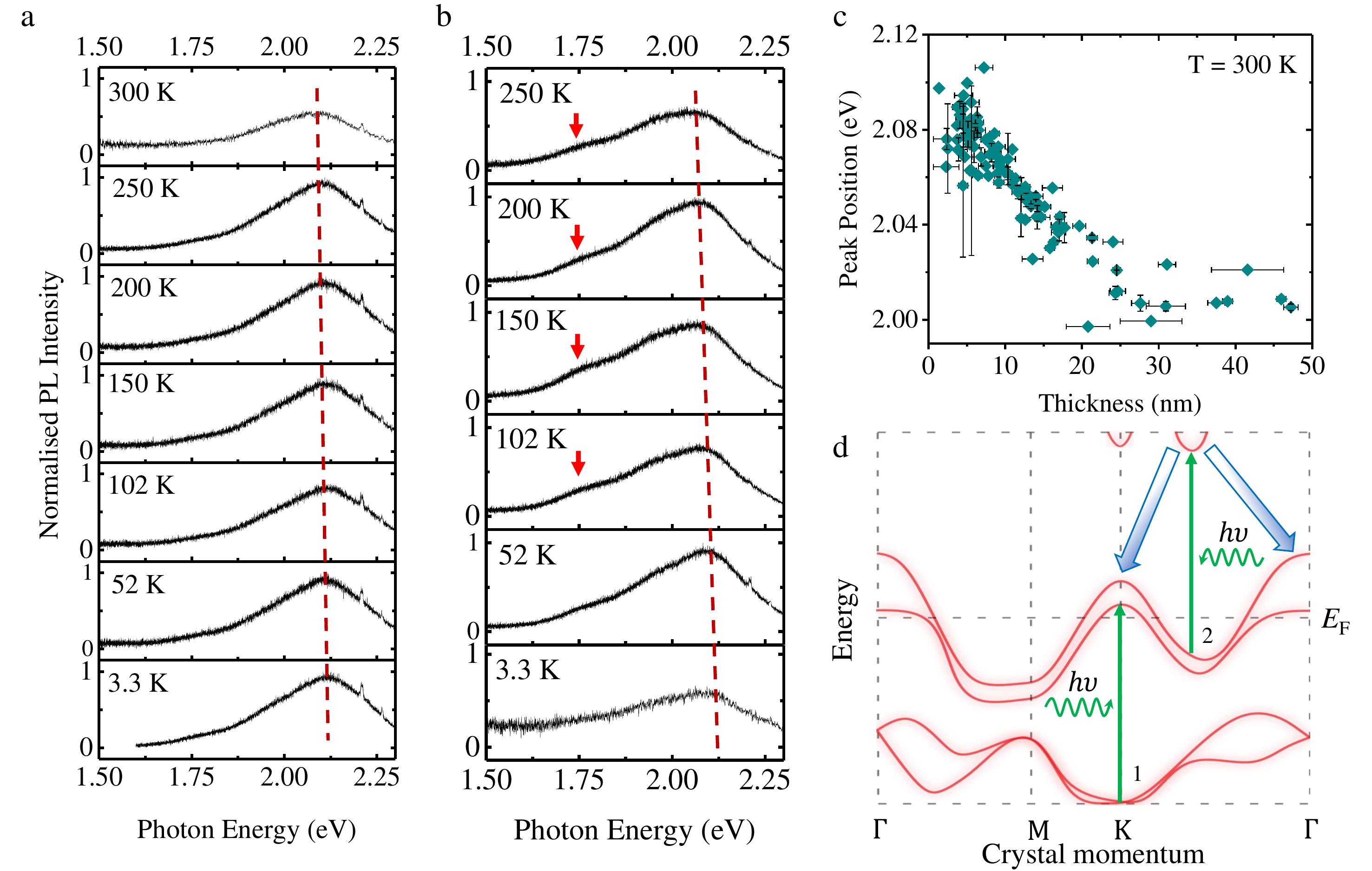}
%\vspace{-1.8in}
\caption{\textbf{Photoluminescence (PL) from 2H-TaSe\tsub2.} (a)-(b) Temperature dependent PL spectra of monolayer [in (a)] and $50$ nm thick [in (b)] 2H-TaSe\tsub2 flakes. (c) Variation of the PL peak position as the function of thickness. The error bars are obtained by characterizing a large number of flakes, and several measurements from different positions on each flake. (d) Schematic bandstructure of 2H-TaSe\tsub2 with crystal momenta along different symmetry directions. The green arrows depict possible inter-band transitions for light absorption, while the blue arrow indicate possible inter-band carrier relaxation paths.}\label{fig:fig2}
\end{figure}
\newpage
\begin{figure}[!hbt]
\centering
%\vs{-0.1in}
%\hs{-1in}
\includegraphics[scale=0.5]{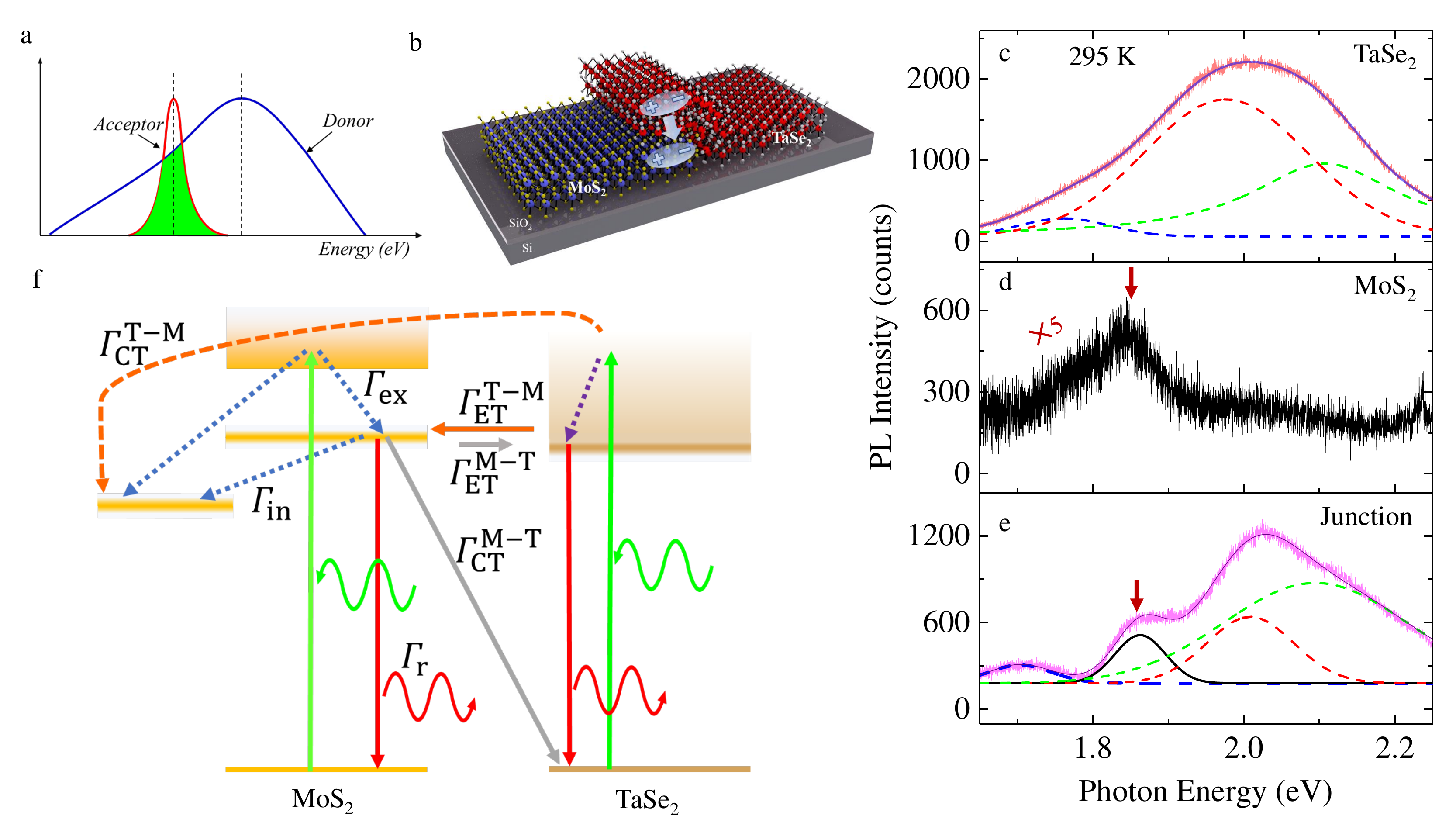}
%\vspace{-1.8in}
\caption{\textbf{Non-radiative resonant energy transfer (NRET) between TaSe\tsub2 and MoS\tsub2.}(a) Schematic representation of spectral overlap required between a donor and an acceptor for NRET process. (b) Schematic representation in-plane dipole-dipole interaction between MoS\tsub2 and TaSe\tsub2. (c-e) Photoluminescence (PL) spectra from  (c) isolated TaSe\tsub2, (d) isolated multi-layer MoS\tsub2, and (e) MoS\tsub2/TaSe\tsub2 junction at $295$ K. The PL intensity from multi-layer MoS\tsub2 is very weak in (d), and is multiplied by 5. In (c) and (e), the dashed traces indicate the fitted Voigt peaks from TaSe\tsub2. The fitted MoS\tsub2 Voigt peak from the junction is distinct and is shown by the solid black trace in (e), which suggests a seven-fold enhancement with respect to the isolated multi-layer MoS\tsub2 owing to non-radiative resonant energy transfer from TaSe\tsub2. The arrow marks indicate the position of the MoS\tsub2 peak in (d) and (e). (f) Schematic representation of the different processes when the MoS\tsub2/TaSe\tsub2 stack is excited with a 532 nm laser.}\label{fig:fig3}
\end{figure}
\newpage
\begin{figure}[!hbt]
\centering
%\vs{-0.1in}
%\hs{-1in}
\includegraphics[scale=0.5]{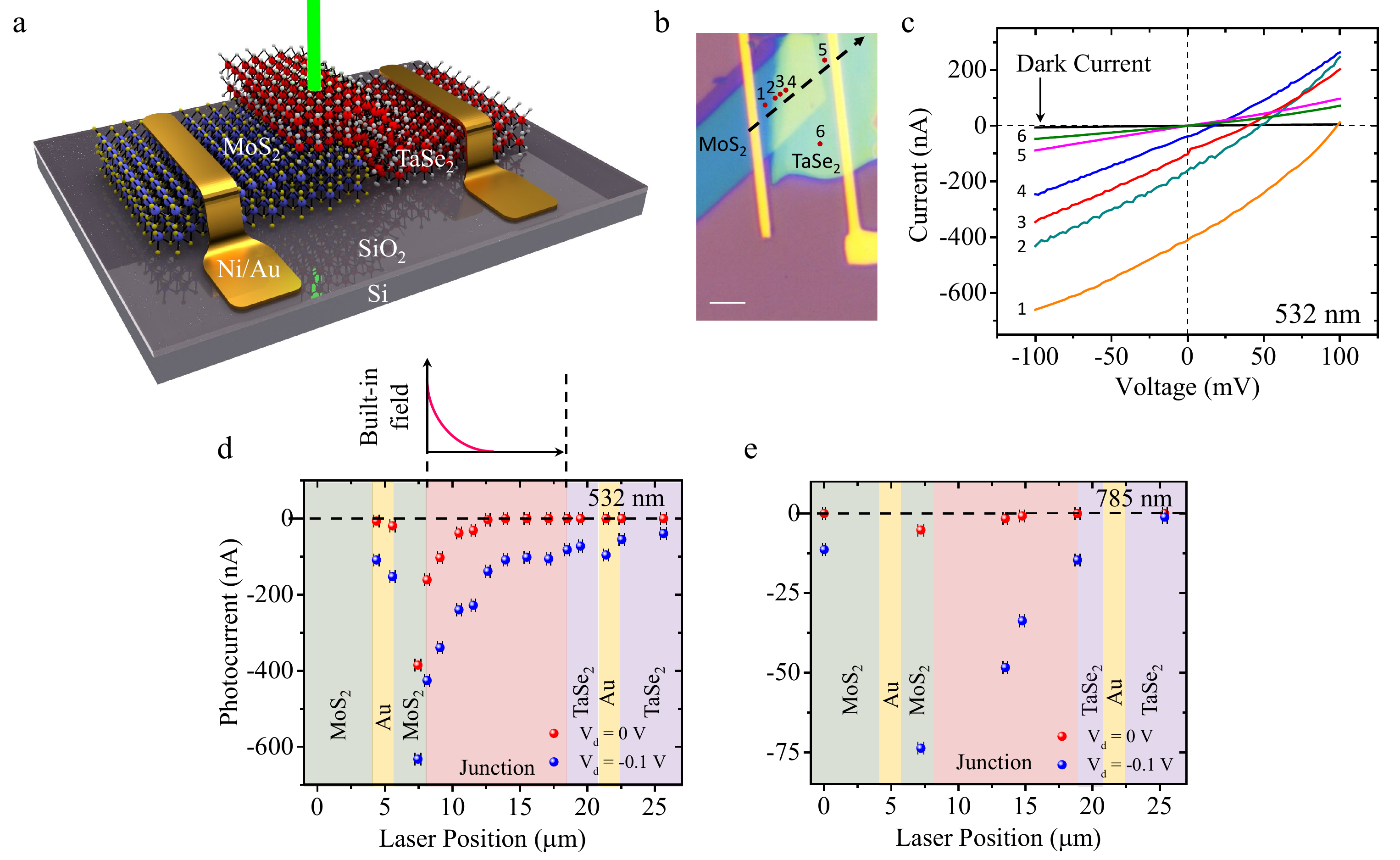}
%\vspace{-1.8in}
\caption{\textbf{Hot Carrier injection from TaSe\tsub2 to MoS\tsub2.} (a) Schematic representation of the device. (b) Optical image of the fabricated heterojunction device indicating laser excitation spots (1-6) for 532 nm wavelength. The arrow indicates the direction of laser scan. Scale bar is $5$ $\mu$m. (c) Current − voltage  characteristics with laser spot position 1-6 as indicated in (b). The dark current from the device is also shown for reference. (d)-(e) Spatial distribution of the photocurrent generated as a function of the laser position for an excitation wavelength of 532 nm [in (d)] and 785 nm [in (e)], at two biasing conditions, 0 V (in red spheres) and -0.1 V (in blue spheres). The error bars indicate the position accuracy of the laser spot. The top panel of (d) schematically shows the spatial distribution of the built-in electric field in the junction region.}\label{fig:fig4}
\end{figure}
\newpage
\begin{figure}[!hbt]
\centering
%\vs{-0.1in}
%\hs{-1in}
\includegraphics[scale=0.5]{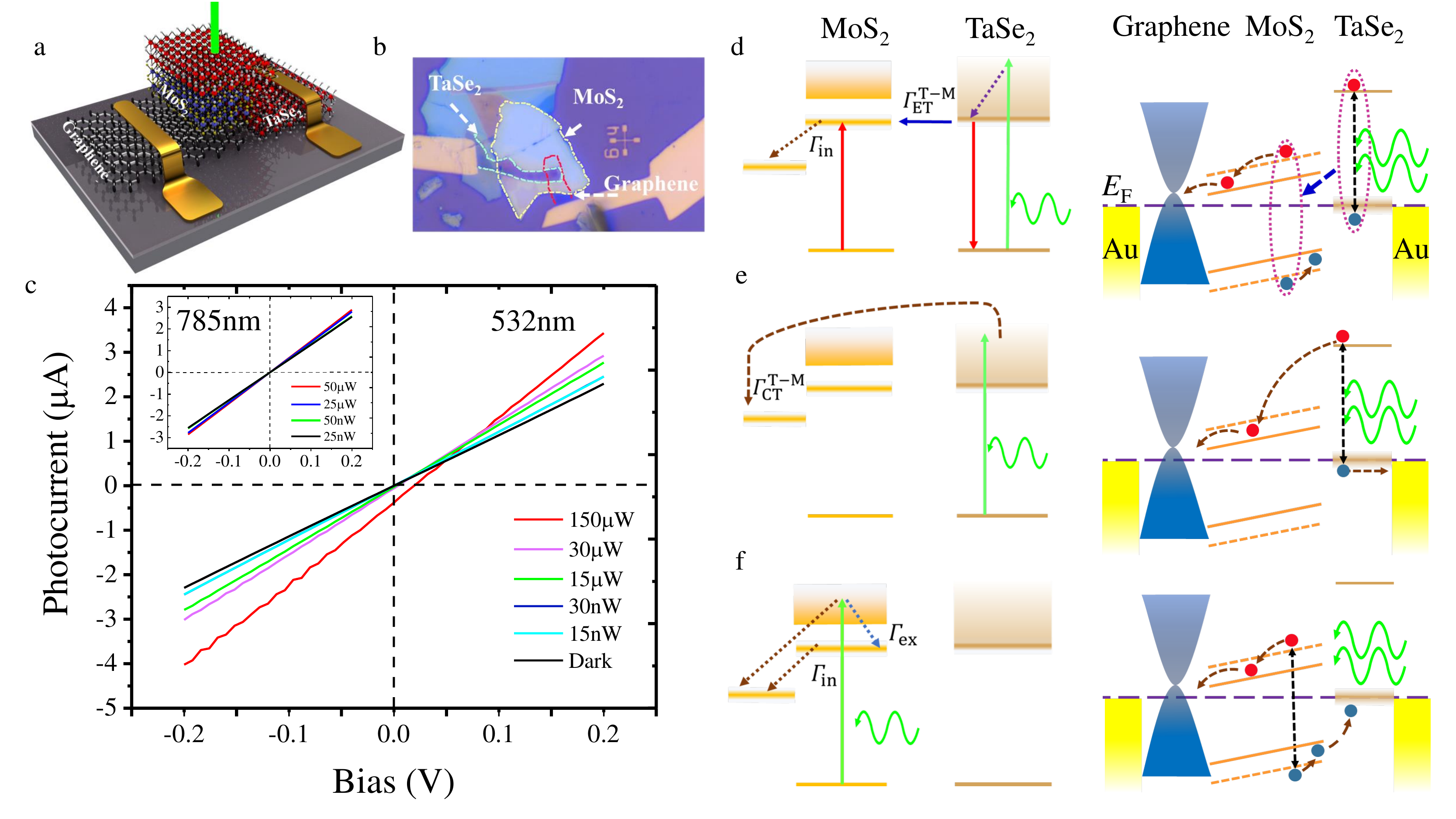}
%\vspace{-1.8in}
\caption{\textbf{Photocurrent generation mechanism in TaSe\tsub2/MoS\tsub2/graphene vertical heterojunction.} (a)-(b) Schematic diagram and optical micrograph of the device. (c) Current-voltage characteristics of the device with various incident power of 532 nm laser. Inset: Current-voltage characteristics with 785 nm laser at different incident powers. (d)-(f) Schematic representation various photocurrent generation mechanisms in the heterojunction. The left panels represent transition diagrams and the right panels indicate the corresponding process in the electronic band diagram. The mechanism in (d) is governed by absorption in TaSe\tsub2 followed by energy transfer in MoS\tsub2 and subsequent dissociation of exciton to lower energy indirect valleys. In (e), light is absorbed by TaSe\tsub2 followed by hot carrier transfer to indirect valley in MoS\tsub2. In (f), the mechanism is governed by direct absorption in MoS\tsub2.}\label{fig:fig5}
\end{figure}
\newpage
\begin{figure}[!hbt]
\centering
%\vs{-0.1in}
%\hs{-1in}
\includegraphics[scale=0.5]{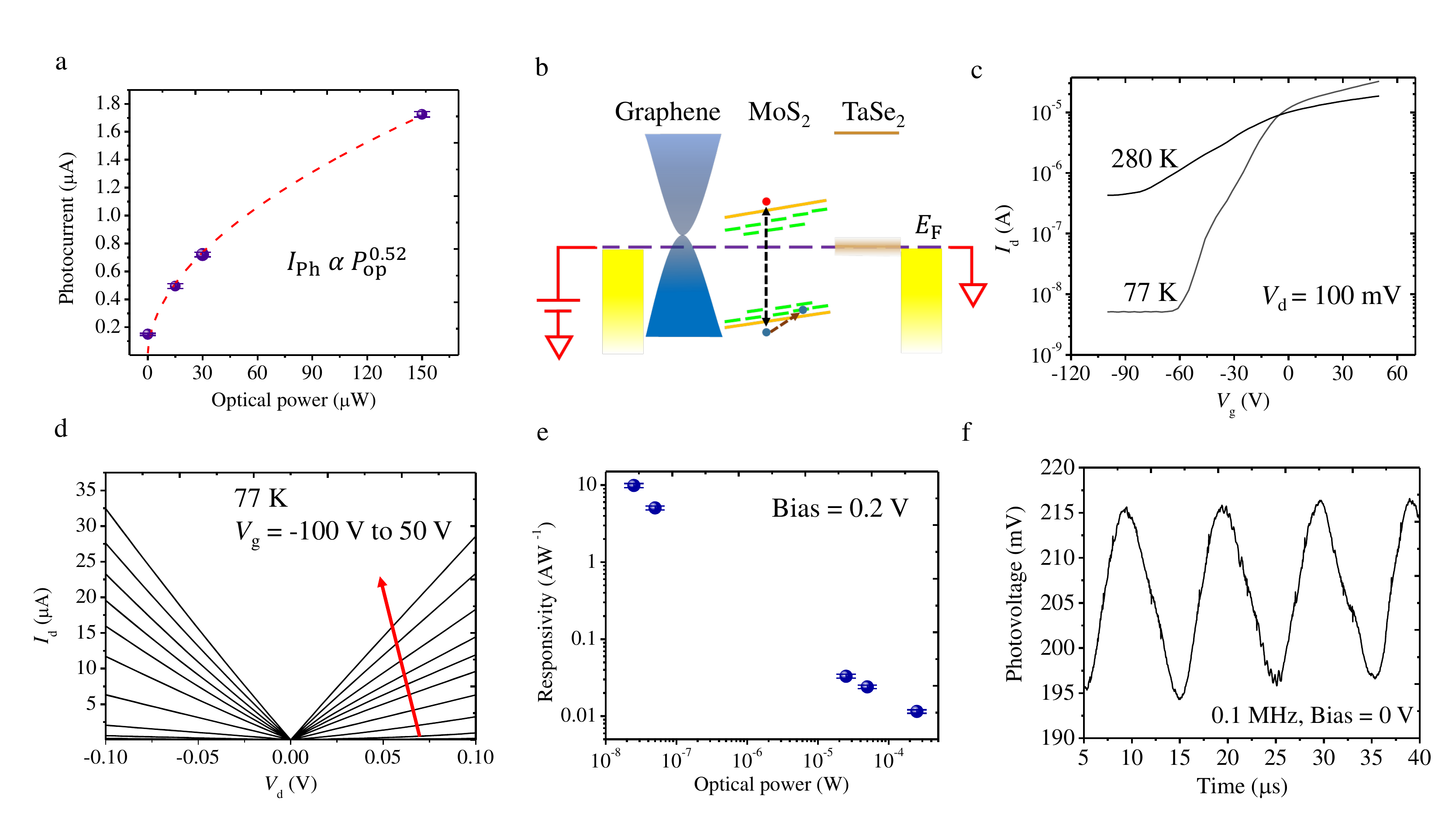}
%\vspace{-1.8in}
\caption{\textbf{Performance of TaSe\tsub2/MoS\tsub2/graphene photodetector.}(a) Photocurrent as a function of incident optical power for 532 nm  wavelength. Exponent in the power law fitting is 0.52, showing the role of traps in the device performance. The error bars are obtained from multiple measurements. (b) Schematic illustration of hole trapping in the existing gap states in MoS\tsub2. (c) Transfer characteristics ($I_\textrm{d}$ versus $V_\textrm{g}$) of a vertical device at both 77K and 280K. Leakage current in the OFF state increases with temperature, indicating trap assisted tunneling through the bandgap of MoS\tsub2. (d) Output characteristics ($I_\textrm{d}$ versus $V_\textrm{d}$) of vertical device at 77K. Gate bias is varied from -100 V to 50 V. (e) Variation of responsivity with incident optical power at a wavelength of 532 nm. Device is biased at 200 mV. The error bars are obtained from multiple measurements. (f) Transient photoresponse of the device measured at zero external bias when the light intensity is modulated at 0.1 MHz.}\label{fig:fig6}
\end{figure}


\section*{Supplementary material}

**Supplementary Note 1**

**Characterization of bulk 2H-TaSe$_2$ crystal**

The bulk 2H-TaSe$_2$ crystals are characterized through scanning electron microscopy (SEM), X-ray diffraction (XRD), and electron-dispersive X-ray spectroscopy (EDS). The SEM image of the layered crystal is shown in Supplementary Figure 1a. Supplementary Figure 1b shows the (002), (006), (008), and (0010) XRD peaks from the bulk crystal, suggesting the high crystalline quality. The elemental analysis of the crystal is performed using EDS in Supplementary Figure 1c-d. The Ta and Se peaks are shown in Supplementary Figure 1c, and the extracted weight and atomic percentages of the two elements in the bulk crystal are shown in the table in Supplementary Figure 1d. A Se:Ta atomic percentage ratio of 2.077 is obtained.

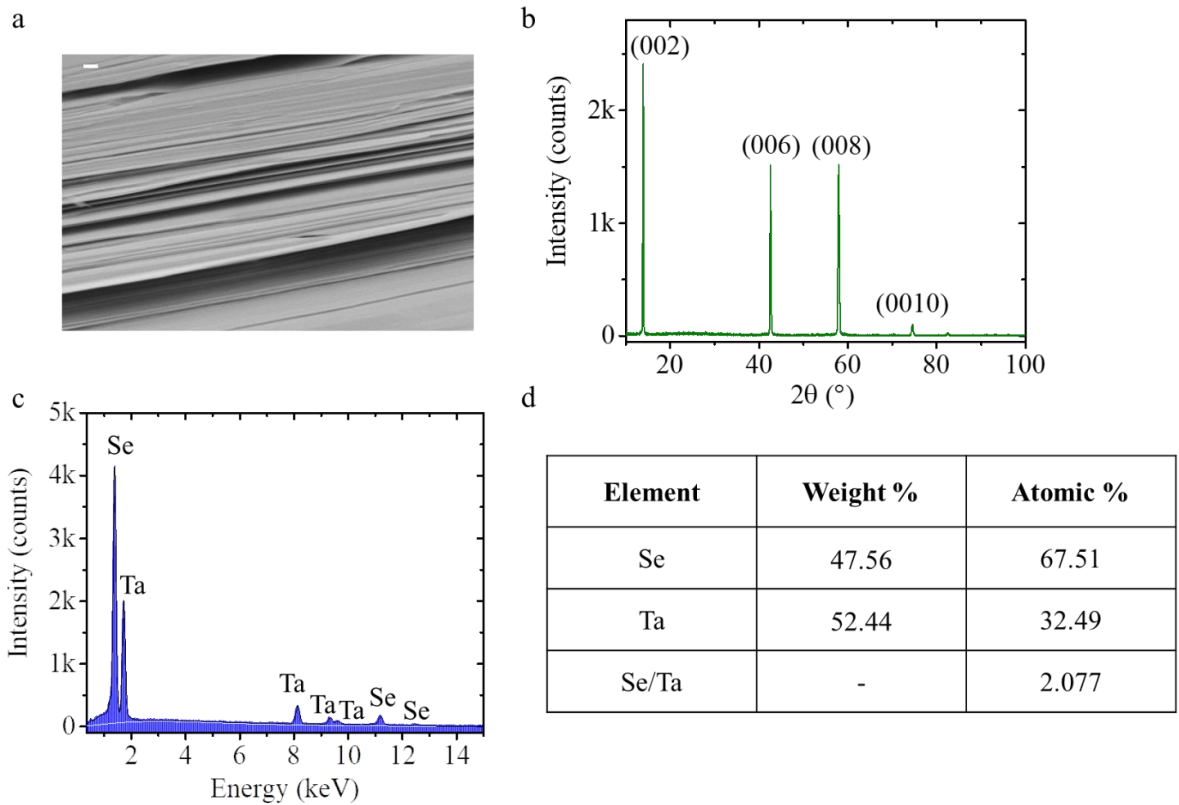

**Supplementary Figure 1: Material Characterization of bulk 2H-TaSe$_2$ crystal**. (a) Scanning electron micrograph of a bulk crystal. Scale bar is 200 nm. (b) X-ray diffraction peaks, and (c-d) elemental analysis using electron-dispersive X-ray spectroscopy of bulk 2H-TaSe$_2$ crystal.

**Supplementary Note 2**

**X-ray Photoelectron Spectroscopy (XPS) of 2H-TaSe$_2$ flakes**

XPS is performed on exfoliated 2H-TaSe$_2$ flakes on Si/SiO$_2$ substrate in an ultra-high vacuum condition. The samples are sputter etched for 30 s before measurements. Supplementary Figure 2a-b show the Ta 4f and Se 3d core level binding energy peaks, which correspond to TaSe$_2$.

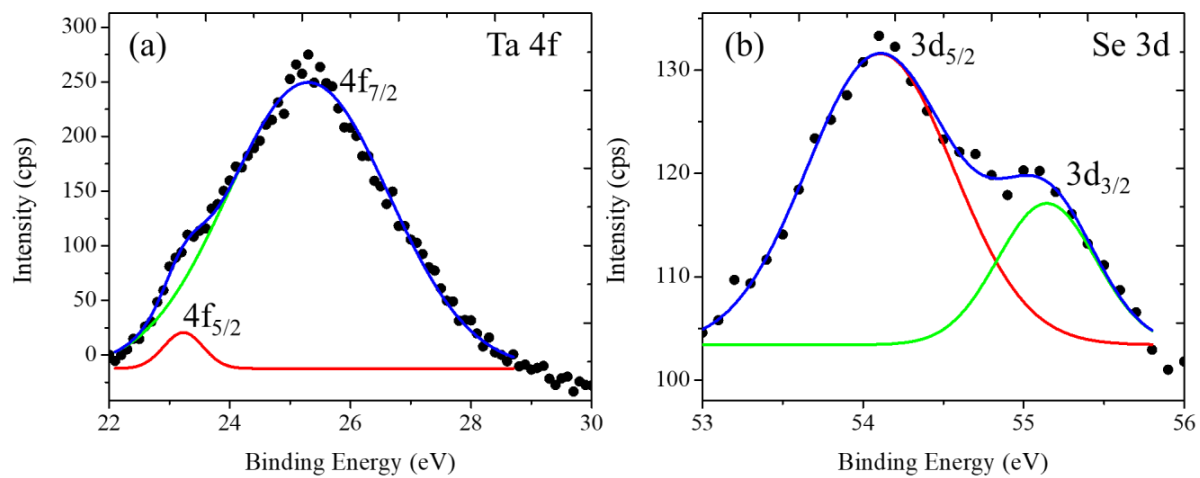

**Supplementary Figure 2: XPS analysis of the 2H-TaSe$_2$ flakes after *in-situ* sputter etch.** Measured XPS spectra of (a) Ta 4f and (b) Se 3d core levels. The black symbols represent the experimental spectra, and the solid lines represent the fitted peaks.

In Supplementary Figure 3, the XPS data without the sputter etch shows a small tantalum oxide shoulder towards the higher energy, suggesting formation of an ultra-thin oxide layer.

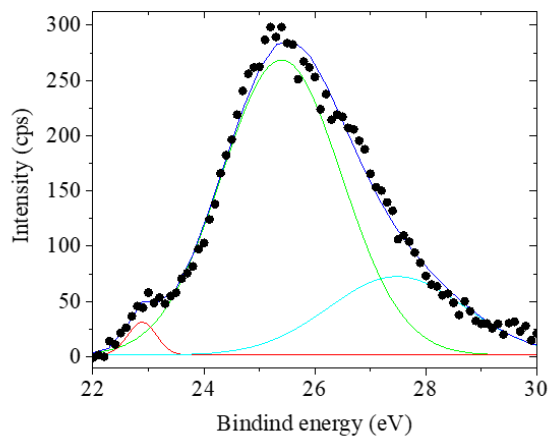

**Supplementary Figure 3: XPS data before *in-situ* sputter etch.** The symbols show experimentally obtained XPS spectrum for Ta 4f core level, with the solid lines indicating fitted peaks. The red and green lines correspond to Ta 4f$_{5/2}$ and Ta 4f$_{7/2}$ peaks, while the light blue peak on the higher energy side indicates presence of an ultra-thin tantalum oxide (TaO$_x$) layer.

**Supplementary Note 3**

**Air stability of 2H-TaSe$_2$ flakes**

**Short term stability:** Supplementary Figure 4 below shows the optical image of an exfoliated flake, where the optical images are captured once every day (day 1 to day 8) in ambient condition. The measurements are performed in ambient condition and the samples are stored in desiccator during the period. The samples do not show any significant visual degradation during the entire period. In the third row of Supplementary Figure 4, we also show an image of the flake depicting thickness mapping obtained using atomic force microscopy on day 12.

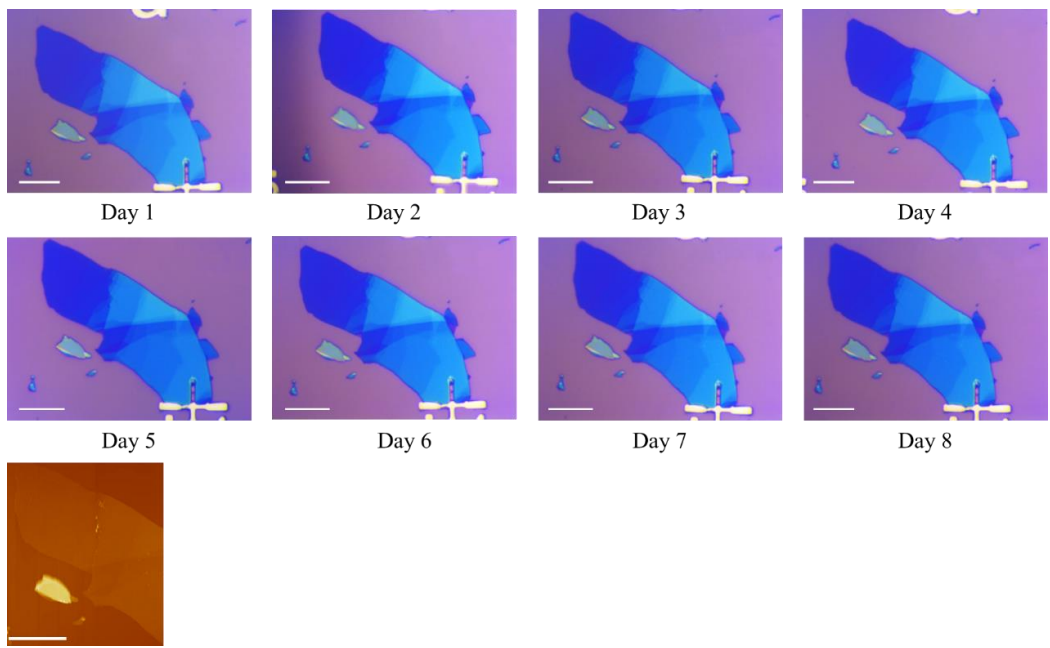

**Supplementary Figure 4: Short term stability of 2H-TaSe$_2$.** Top and middle panel: Optical image of an exfoliated flake from day 1 to day 8. Bottom panel: Thickness mapping using Atomic Force Microscopy (AFM) image of the same flake taken on day 12. Scale bar is 20 μm.

**Laser induced oxidation:** We further perform intentional oxidation of a flake by laser induced heating, and study the evolution of the Raman peak from the TaSe$_2$ flake. It was reported earlier[1] that such laser induced oxidation causes appearance of a new strong Raman peak around 255 cm$^{-1}$. Supplementary Figure 5 shows the optical image and the Atomic Force Microscopy (AFM) image of the flake under investigation. The spot shown by white arrow in Supplementary Figure 5b is obtained after intentional laser induced oxidation by exposing with a focused laser beam of few mW. The Raman spectra for the flake (before and after the laser burn) clearly show that a strong peak around 255 cm$^{-1}$ appears (which is stronger than the TaSe$_2$ peak around 236 cm$^{-1}$) due to the laser oxidation, which is otherwise completely absent. This is in good agreement with the previous report[1] and suggests that the TaSe$_2$ flakes are quite stable in ambience unless such intentional oxidation is performed.

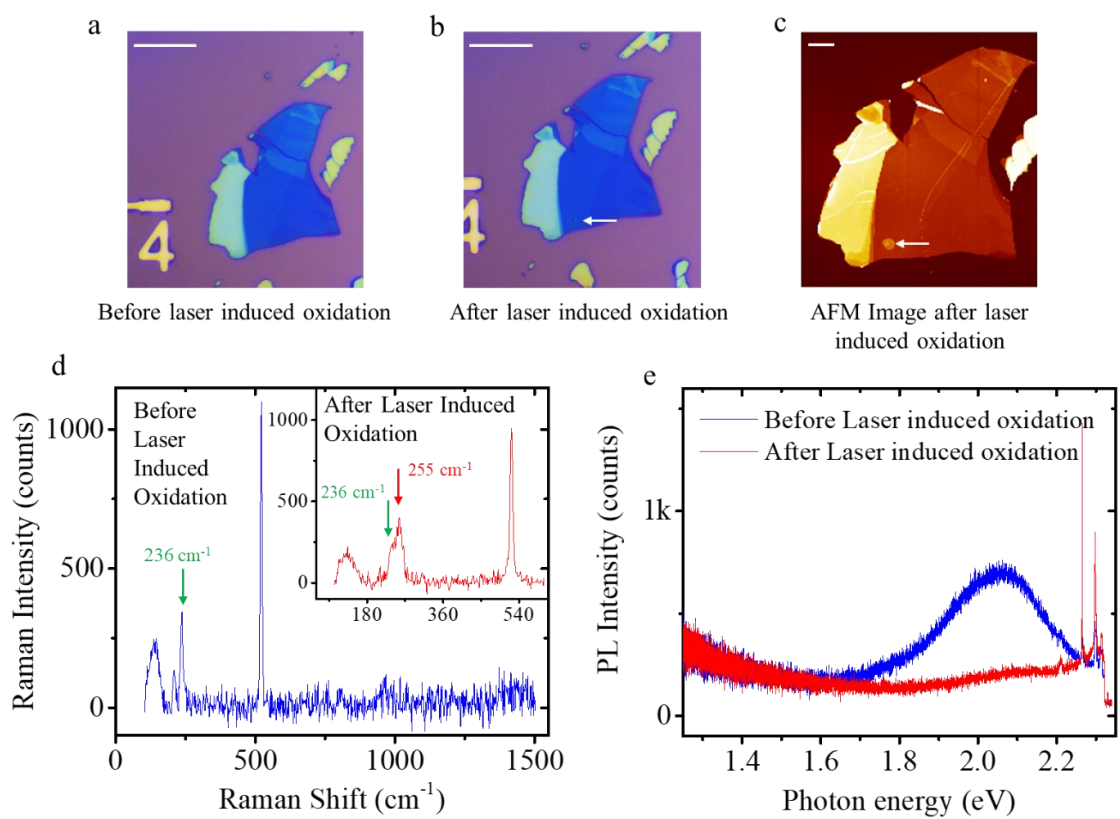

**Supplementary Figure 5: Laser induced oxidation of 2H-TaSe$_2$.** (a) Optical image of the flake before laser induced oxidation (Scale bar: 20 μm). (b) Laser burn performed at the spot shown by white arrow (Scale bar: 20 μm). (c) AFM image of the flake with the laser burn spot shown by white arrow (Scale bar: 5 μm). (d) Raman signal showing the 2H-TaSe$_2$ peaks of the original sample. Inset: Raman data taken from the laser burn spot, showing appearance of strong peak around 255 cm$^{-1}$. (e) The photoluminescence obtained from the oxidized portion (in red), indicating weak intensity.


**Effect of device fabrication and long term (12 months) stability of devices:** Finally, to understand the effect of device processing steps and long term air stability, we perform optical imaging and Raman study (see Supplementary Figure 6) for the device which was fabricated 12 months earlier, has gone through several measurement cycles over a week in ambient condition and then stored in a desiccator. Again, we do not clearly observe any visual degradation of the device through optical images even after 12 months. The Raman data, taken after 12 months, still shows strong TaSe$_2$ vibrational peaks, clearly indicating sample integrity, though the background noise is slightly larger than the fresh samples which is usual for other layered-materials as well. There appears to be a very weak peak around 255 cm$^{-1}$, although is almost buried in the background noise (and much weaker than the TaSe$_2$ 236 cm$^{-1}$ peak) – which suggests possible surface oxidation. This is the combined effect of device fabrication and 12 months of exposure of the device to ambience.

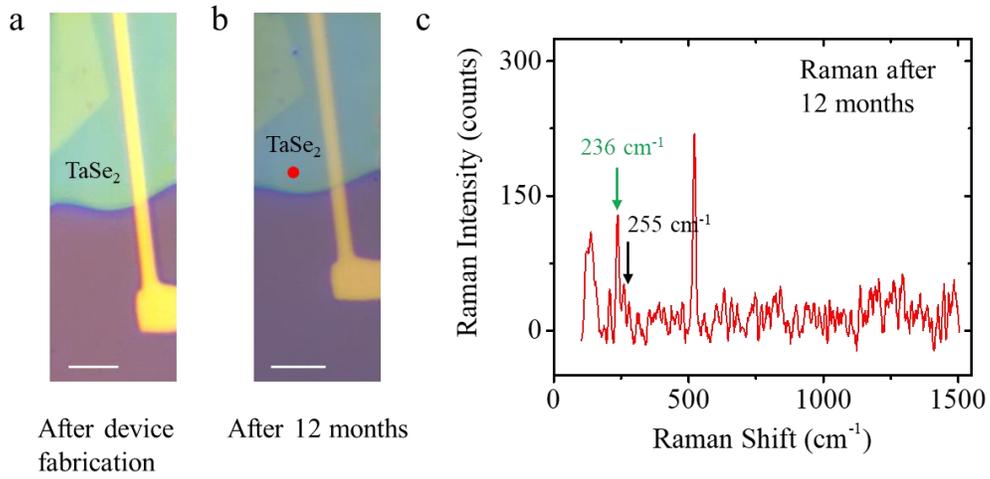

**Supplementary Figure 6: Long term stability of 2H-TaSe$_2$.** (a) Optical image after device fabrication (Scale bar: 5 μm). (b) Optical image after 12 months (Scale bar: 5 μm). (c) Raman spectrum of the TaSe$_2$ after 12 months. The laser spot is shown in red spot in (b).

**Supplementary Note 4**

We obtain the photoluminescence spectrum from 2H-TaSe$_2$ in a wide spectral range in order to simultaneously show the Raman lines and the photoluminescence peak, as indicated in Supplementary Figure 7a. The Raman lines are shown in a zoomed-in range in the right inset of the figure in blue, red, and black arrows. Separately, the Raman spectrum is obtained and shown in Supplementary Figure 7b, with the corresponding peaks are marked by the arrows of the same colour. The Raman peaks indicate the absence of any effect from a possible surface-oxide layer, which in turn suggests that the photoluminescence arises from the bulk of 2H-TaSe$_2$, and not from any surface-oxide layer.

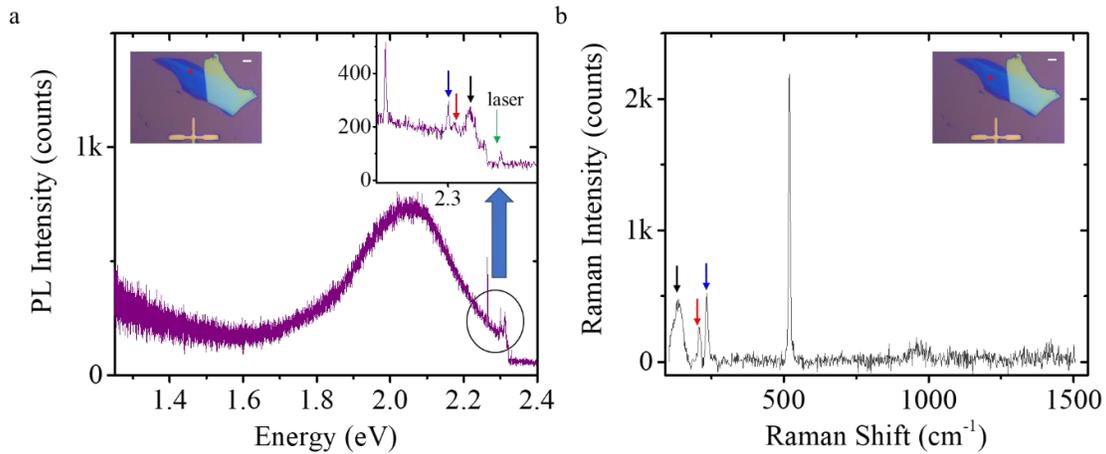

**Supplementary Figure 7: Full spectrum of 2H-TaSe$_2$ photoluminescence.** (a) Full spectrum with the entire detector range (from 1.25 eV to cut-off of the edge filter), showing simultaneously the PL feature and the sharp Raman lines close to the excitation. Left inset: Optical image of the flake with the red dot indicating the laser spot. Scale bar is 5 μm. Right inset: The portion of the spectrum with Raman lines is shown in a zoomed-in view. The different Raman peaks are colour coded in black, red and blue, in increasing values of Raman shift. The excitation laser line is indicated by the green arrow. (b) The corresponding peaks in Raman measurement (in cm$^{-1}$). No tantalum oxide Raman peak is observed during the PL measurement. Inset: Optical image of the flake with the red dot indicating the laser spot. Scale bar is 5 μm.

**Supplementary Note 5**

**2H-TaSe$_2$ photoluminescence after exfoliation in inert atmosphere**

To establish the point that any ambience induced surface oxidation (during exfoliation in ambient condition) does not give rise to the observed photoluminescence, we perform exfoliation followed by Poly(methyl methacrylate) (PMMA) spin coating inside a glovebox with N$_2$ atmosphere. Photoluminescence and Raman experiment are then performed on the quoted samples using a laser power of 40 µW, as shown below. The coated samples exhibit similar photoluminescence and Raman characteristics as the ambient exfoliated samples. The relative thickness of the PMMA coated samples are judged by the Raman intensity of the Si peak, with the thickest TaSe$_2$ sample to exhibit weakest Si Raman intensity.

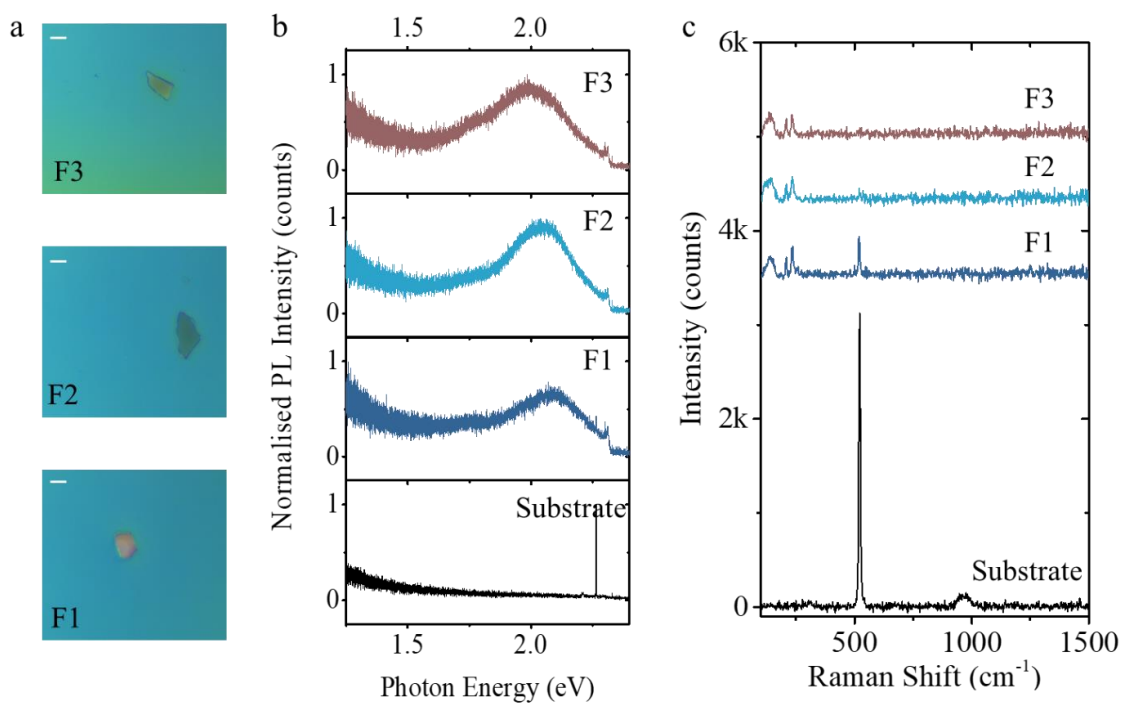

**Supplementary Figure 8: Characteristics of flakes exfoliated in inert atmosphere.** (a) Optical images of the PMMA quoted samples. Scale bar is 5 µm. (b) Photoluminescence (PL) spectra of the samples. With an increase in the flake thickness, the PL peak is found to be red shifted. (c) The Raman intensity of the samples, with the Si Raman at ~520.5 cm$^{-1}$ allowing us to estimate the relative thickness of the flakes.

**Supplementary Note 6**

**Thickness dependent differential reflectance spectra of 2H-TaSe$_2$**

Differential reflectance of 2H-TaSe$_2$ flakes with varying thickness is measured using a broadband source illumination, at 295 K. We plot $\frac{\Delta R}{R_0}$ where $\Delta R = R - R_0$, $R$ and $R_0$ being the sample with substrate and only substrate reflectance, respectively. Here the substrate is Si covered with 285 nm SiO$_2$. The broad dip primarily arises due to substrate induced interference effects, and does not show any sharp excitonic feature.

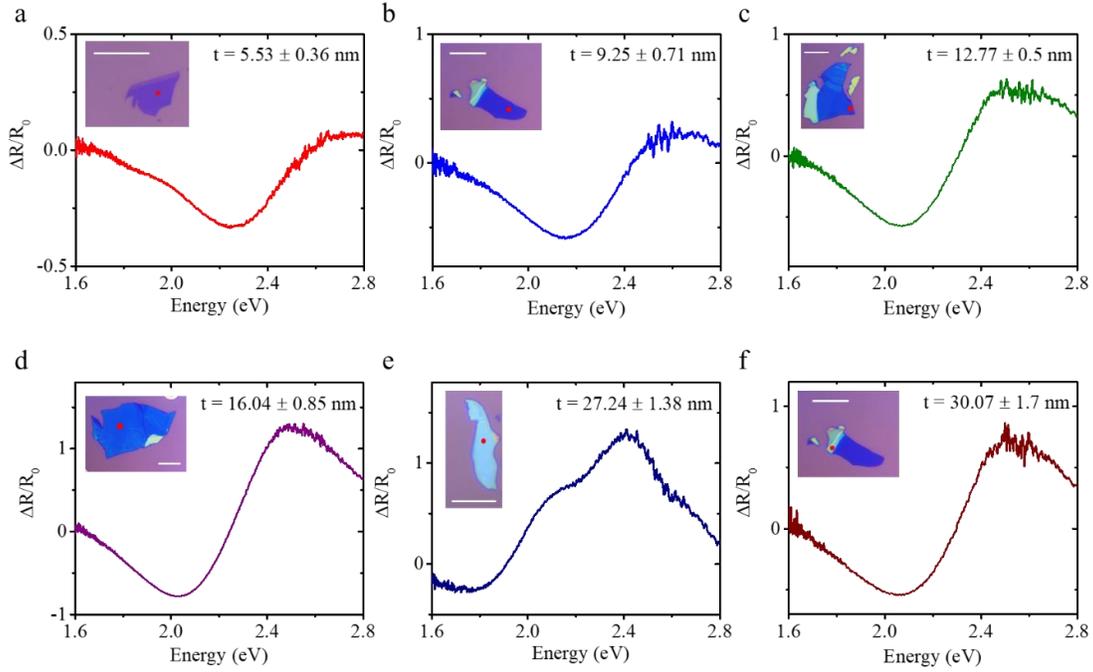

**Supplementary Figure 9: Differential reflectance plot of 2H-TaSe$_2$.** (a-f) Plot of differential reflectance of TaSe$_2$ samples with varying thickness as a function of photon energy. The thickness of the respective samples is given in each figure. The insets show the optical images of the respective flakes with the red dots indicating the position of the laser spot. Scale bar in each inset is 20 μm.

**Supplementary Note 7**

**Time-resolved photoluminescence measurement on 2H-TaSe$_2$**

We illuminate TaSe$_2$ flakes with 50 ps laser pulses at 405 nm wavelength and 40 MHz repetition rate (Picoquant PDL 828 "Sepia II", LDH-D-C-405). The excitation spot diameter is about 4 µm. We collect the photoluminescence signal using an objective lens of numerical aperture 0.9. The collected signal is split between an imaging camera and a photon counting module (Picoquant PMA Hybrid 40 and Picoharp 300) for lifetime measurements. We use a long-pass dichroic mirror with a cutoff wavelength of 550 nm (Thorlabs DMLP550) and a long-pass filter with a cutoff wavelength of 500 nm (Thorlabs FEL0500) to reject any laser reflections.

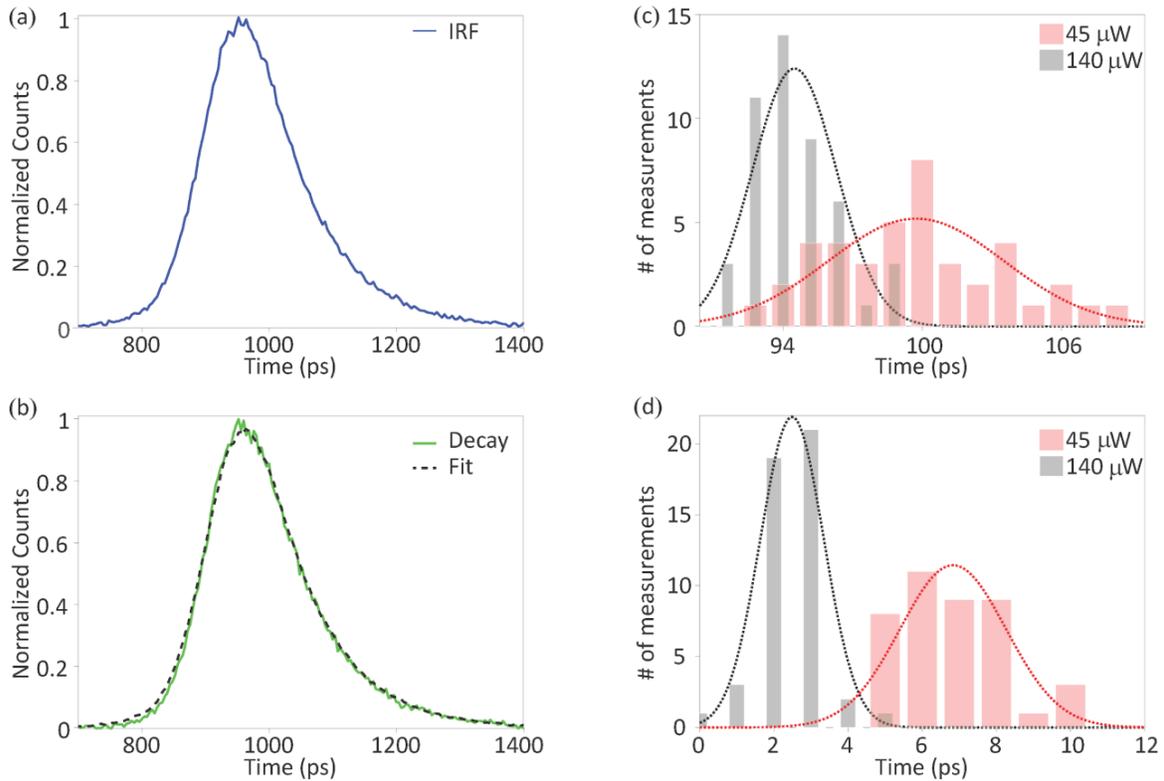

**Supplementary Figure 10: Time-resolved photoluminescence measurement on 2H-TaSe$_2$.** (a) Instrument response function (IRF) measured at the excitation wavelength. (b) Time-resolved photoluminescence intensity of TaSe$_2$ when excited with 45 µW. (c) Distribution of observed lifetimes obtained from single exponential fitting (without deconvolution with IRF), and (d) Distribution of lifetimes extracted after deconvolving the observed data with IRF. The dotted curves in (c) and (d) represent Gaussian fits to the distributions.

Supplementary Figure 10a shows instrument response function (IRF) measured at the excitation wavelength and supplementary Figure 10b shows representative data for time-resolved photoluminescence intensity of TaSe$_2$. The average excitation power is 45 µW. We fit the measured data with a single exponential function (not shown here). The fitted lifetime is 102 ps. After deconvolving the measured data with the instrument response function[2], the

extracted lifetime is obtained as 7.5 ps (fit shown in Supplementary Figure 10b). We Note that the photoluminescence signal from the substrate (SiO$_2$ on Si) is about two orders of magnitude lesser than from TaSe$_2$ under the same experimental conditions.

We perform several measurements across seven TaSe$_2$ flakes for two different excitation powers (45 µW and 140 µW). Supplementary Figure 10c shows observed lifetimes obtained from single exponential fitting (without deconvolution with IRF) across all measurements. The dotted curves represent Gaussian fit to the distributions. The mean values of the observed lifetimes are 99.8 ps for 45 µW and 95.4 ps for 140 µW of excitation power. These observed lifetimes are close to the full width at half maximum of IRF (~ 95 ps) of our measurement setup. Therefore, the actual lifetimes may be shorter than these observed values.

Supplementary Figure 10d shows lifetimes extracted after deconvolving the observed data with IRF across all measurements. The dotted curves represent Gaussian fit to the distributions. The mean values of extracted lifetimes are 6.8 ps for 45 µW and 2.5 ps for 140 µW of excitation power. We Note that deconvolution can accurately extract lifetimes only up to 10% of the IRF width[3]. Hence, while there might be some inaccuracy in these extracted values, this analysis suggests that photoluminescence decay lifetimes in TaSe$_2$ are shorter than 10 ps. This observation is further supported by small standard deviation values in the distributions (1.4 ps for 45 µW and 0.9 ps for 140 µW of excitation power).

**Supplementary Note 8**

**Rate equations for energy transfer**

$$\frac{dN_{ex}^{M,jun}}{dt} = \Gamma_{ex} N_{e-h}^{M,jun} - (\Gamma_d + \Gamma_{CT}^{M-T} + \Gamma_{ET}^{M-T}) N_{ex}^{M,jun} + \Gamma_{ET}^{T-M} N_{e-h}^{T,jun} \quad (1)$$

$$\frac{dN_{e-h}^{T,jun}}{dt} = G_f^{T,jun} - (\Gamma_s + \Gamma_{ET}^{T-M}) N_{e-h}^{T,jun} + (\Gamma_{CT}^{M-T} + \Gamma_{ET}^{M-T}) N_{ex}^{M,jun} \quad (2)$$

where $G_f^{T,jun}$ is the effective generation rate of e-h pairs in the energy state of TaSe$_2$ that is resonant to the exciton state in MoS$_2$ and $\Gamma_s$ is the scattering rate of these e-h pairs from this state to other non-resonant energy states in TaSe$_2$. $\Gamma_d = \Gamma_r + \Gamma_{nr}$ is the combined radiative and non-radiative decay rate of excitons within MoS$_2$ and is assumed to be similar on SiO$_2$ and TaSe$_2$.

With the observation that the photoluminescence of A$_{1s}$ peak of MoS$_2$ on TaSe$_2$ enhances significantly relative to that of MoS$_2$ on SiO$_2$, the first term in equation (1) representing the direct absorption of MoS$_2$ can be dropped out. The steady state photoluminescence intensity of A$_{1s}$ peak is then given by

$$I_M^{jun} = \Gamma_r N_{ex}^{M,jun} = \frac{G_{e-h}^{T,jun} \Gamma_{ET}^{T-M} \left[\frac{\Gamma_r}{\Gamma_d}\right]}{\Gamma_s \left[1 + \frac{\Gamma_{CT}^{M-T} + \Gamma_{ET}^{M-T}}{\Gamma_d}\right] + \Gamma_{ET}^{T-M}} \quad (3)$$

The PL enhancement factor (α) is defined to be the ratio of steady state PL intensity of MoS$_2$ on TaSe$_2$ to that of on SiO$_2$ and turns out to be

$$\alpha = \frac{I_M^{jun}}{I_M} = \frac{G_{e-h}^{T,jun}}{G_{ex}^M} \left\{ \frac{\Gamma_{ET}^{T-M}}{\Gamma_s \left[1 + \frac{\Gamma_{CT}^{M-T} + \Gamma_{ET}^{M-T}}{\Gamma_d}\right] + \Gamma_{ET}^{T-M}} \right\} \quad (4)$$

## Supplementary Table 1

**Performance comparison of 2D-material based photodetectors[4-14] (with response time < 100 ms)**

| Material stack | Architecture | Bias (V) | Wavelength (nm) | Responsivity (AW$^{-1}$) | Response time | Ref. |
|---|---|---|---|---|---|---|
| **TaSe$_2$/MoS$_2$/Graphene** | **Vertical** | **0** | **532** | **10** | **4 µs** | **This work** |
| GaAs/MoS$_2$ | Vertical | 0 | 635 | 0.321 | 17 µs | 4 |
| MoS$_2$/Graphene/WS$_2$ | Vertical | 1 | 532 | 10$^4$ | 35 µs | 5 |
| Graphene/MoTe$_2$/Graphene | Vertical | 0 | 1064 | 0.11 | 24 µs | 6 |
| Graphene/InSe/Graphene | Lateral | 1 | 633 | 4x10$^4$ | 1 ms | 7 |
| Graphene/MoS$_2$/Graphene | Vertical | 0 | 488 | 0.22 | 50 µs | 8 |
| MoS$_2$/Si | Vertical | 0 | 808 | 0.1 | 3 µs | 9 |
| MoS$_2$ | Lateral | 5 | 635 | 5x10$^4$ | 8 ms | 10 |
| MoS$_2$/Si | Vertical | 2 | 808 | 1 | 56 ns | 11 |
| MoTe$_2$/MoS$_2$ | Lateral | 0.5 | 637 | 0.046 | 60 µs | 12 |
| BP | Lateral | 0.2 | 640 | 0.0048 | 1 ms | 13 |
| ITO/MoS$_2$/Cu$_2$O | Vertical | 0.5 | 532 | 5.77x10$^4$ | < 70 ms | 14 |


\end{document}